\documentclass[fleqn,usenatbib]{mnras}

\usepackage{newtxtext,newtxmath}

\usepackage[T1]{fontenc}

\DeclareRobustCommand{\VAN}[3]{#2}
\let\VANthebibliography\thebibliography
\def\thebibliography{\DeclareRobustCommand{\VAN}[3]{##3}\VANthebibliography}

\usepackage{graphicx}	
\usepackage{amsmath}	
\usepackage{caption}
\usepackage{subcaption}
\usepackage{tikz,xcolor,hyperref}

\definecolor{lime}{HTML}{A6CE39}
\DeclareRobustCommand{\orcidicon}{%
    \begin{tikzpicture}
    \draw[lime, fill=lime] (0,0) 
    circle [radius=0.16] 
    node[white] {{\fontfamily{qag}\selectfont \tiny ID}};
    \draw[white, fill=white] (-0.0625,0.095) 
    circle [radius=0.007];
    \end{tikzpicture}
    \hspace{-2mm}
}

\newcommand{\orcidSP}{\href{https://orcid.org/0000-0001-5579-9487}{\orcidicon}}
\newcommand{\orcidRW}{\href{https://orcid.org/0000-0002-6122-7052}{\orcidicon}}
\newcommand{\orcidTB}{\href{https://orcid.org/0000-0002-2853-0834}{\orcidicon}}
\newcommand{\orcidJO}{\href{https://orcid.org/0000-0002-9079-1883}{\orcidicon}}
\newcommand{\orcidOF}{\href{https://orcid.org/0000-0002-8985-4277}{\orcidicon}}
\newcommand{\orcidDL}{\href{https://orcid.org/0000-0002-6894-6172}{\orcidicon}}

\usepackage{acronym}

\acrodef{PUC}[PUC]{Pontifical Catholic University of Chile}
\acrodef{INT}[INT]{Isaac Newton Telescope}
\acrodef{WHT}[WHT]{William Herschel Telescope}
\acrodef{ING}[ING]{Isaac Newton Group of Telescopes}
\acrodef{GMT}[GMT]{Giant Magellan Telescope}

\acrodef{SLODAR}[SLODAR]{SLOpe Detection and Ranging}
\acrodef{SL-SLODAR}[SL-SLODAR]{Surface Layer SLODAR}
\acrodef{RoboDIMM}[RoboDIMM]{Robotic Differential Image Motion Monitor}
\acrodef{SCIDAR}[SCIDAR]{SCIntillation Detection And Ranging}
\acrodef{Stereo-SCIDAR}[Stereo--SCIDAR]{Stereo--SCIntillation Detection And Ranging}
\acrodef{LOLAS}[LOLAS]{LOw LAyer SCIDAR}
\acrodef{CO-SLIDAR}[CO-SLIDAR]{COupled SLodar scIDAR}
\acrodef{SCO-SLIDAR}[SCO-SLIDAR]{Single COupled SLodar scIDAR}
\acrodef{LLA-SCIDAR}[LLA-SCIDAR]{LensLet Array SCIDAR}
\acrodef{ESO}[ESO]{European Southern Observatory}
\acrodef{AT}[AT]{Auxillary Telescope}
\acrodef{VLT}[VLT]{Very Large Telescope}
\acrodef{GSM}[GSM]{Generalized Seeing Monitor}
\acrodef{FASS-SHIMM}[FASS-SHIMM]{Full Aperture Scintillation Sensor - Shack Hartmann Image Motion Monitor}
\acrodef{FASS}[FASS]{Full Aperture Seeing Sensor}
\acrodef{MASS}[MASS]{Multi Aperture Scintillation Sensor}
\acrodef{DIMM}[DIMM]{Differential Image Motion Monitor}
\acrodef{SHIMM}[SHIMM]{Shack Hartmann Image Motion Monitor}
\acrodef{SH}[SH]{Shack Hartmann}
\acrodef{WFS}[WFS]{wavefront sensor}
\acrodef{SHWFS}[SHWFS]{Shack--Hartmann wavefront sensor}
\acrodef{AO}[AO]{adaptive optics}
\acrodef{H-DIMM}[H-DIMM]{Hartmann DIMM}
\acrodef{GDIMM}[GDIMM]{Generalized DIMM}
\acrodef{DIMMWIT}[DIMMWIT]{DIMM Which Is Transportable}
\acrodef{FADE}[FADE]{FAst DEfocus monitor}

\acrodef{PMT}[PMT]{Photo-Multipler Tube}
\acrodef{CCD}[CCD]{Charge-Coupled Device}
\acrodef{EMCCD}[EMCCD]{Electron-Multiplying CCD}
\acrodef{PSF}[PSF]{point spread function}
\acrodef{FWHM}[FWHM]{full width half maximum}
\acrodef{Cn2}[J]{total integrated turbulence strength}
\acrodef{L0}[$L_{0}$]{outer scale}
\acrodef{l0}[$l_{0}$]{inner scale}
\acrodef{tau}[$\tau_{0}$]{coherence time}
\acrodef{r0}[$r_{0}$]{Fried parameter}
\acrodef{veff}[$v_{\mathrm{eff}}$]{effective wind-blown turbulence velocity}
\acrodef{mag}[$V$]{V-band magnitude}
\acrodef{SNR}{signal-to-noise ratio}
\acrodef{theta}[$\theta_{0}$]{isoplanatic angle}
\acrodef{dh}[$\delta h$]{vertical resolution}
\acrodef{Hmax}[$h_{\mathrm{max}}$]{maximum altitude}

\acrodef{OTP}[OTP]{optical turbulence profile}

\acrodef{d}[$d$]{sub--aperture width}
\acrodef{covim}[$A^{m}_{\delta i, \delta j}$]{measured spatial auto-covariances}
\acrodef{covit}[$A^{t}_{\delta i, \delta j}$]{theoretical spatial auto-covariance}
\acrodef{theta_travel}[$\theta_{trav}$]{direction of the turbulent layer traversing the telescope aperture with respect to the \ac{SH} lenslet array}

\acrodef{knee}[$f_{\mathrm{knee}}$]{knee frequency}
\acrodef{peakf}[$f_{\mathrm{peak}}$]{peak frequency}

\title[SHIMM: A Versatile Seeing Monitor for Astronomy]{SHIMM: A Versatile Seeing Monitor for Astronomy}

\author[S. Perera et al.]{
Saavidra Perera,$^{1,2}$\thanks{E-mail: sperera@ucsd.edu}\orcidSP
Richard W. Wilson,$^{1}$ \orcidRW
Tim Butterley,$^{1}$ \orcidTB
James Osborn,$^{1}$ \orcidJO
Ollie J. D. Farley, $^{1}$ \orcidOF
\newauthor and Douglas J. Laidlaw $^{1}$ \orcidDL
\\
\\
$^{1}$Department of Physics, Centre for Advanced Instrumentation, Durham University, South Road, Durham, DH1 3LE, UK\\
$^{2}$Center for Astrophysics and Space Sciences, University of California San Diego, 9500 Gilman Dr, La Jolla, CA 92093, USA
}

\date{Accepted XXX. Received YYY; in original form ZZZ}

\pubyear{2022}

\begin{document}
\label{firstpage}
\pagerange{\pageref{firstpage}--\pageref{lastpage}}
\maketitle

\begin{abstract}
Characterisation of atmospheric optical turbulence is crucial for the design and operation of modern ground--based optical telescopes. In particular, the effective application of adaptive optics correction on large and extremely large telescopes relies on a detailed knowledge of the prevailing atmospheric conditions, including the vertical profile of the optical turbulence strength and the atmospheric coherence timescale. The Differential Image Motion Monitor (DIMM) has been employed as a facility seeing monitor at many astronomical observing sites across the world for several decades, providing a reliable estimate of the seeing angle. Here we present the Shack--Hartmann Image Motion Monitor (SHIMM), which is a development of the DIMM instrument, in that it exploits differential image motion measurements of bright target stars. However, the SHIMM employs a Shack-Hartmann wavefront sensor in place of the two--hole aperture mask utilised by the DIMM. This allows the SHIMM to provide an estimate of the seeing, unbiased by shot noise or scintillation effects. The SHIMM also produces a low--resolution (three--layer) measure of the vertical turbulence profile, as well as an estimate of the coherence timescale. The SHIMM is designed as a low-cost, portable, instrument. It is comprised of off-the-shelf components so that it is easy to duplicate and well--suited for comparisons of atmospheric conditions within and between different observing sites. Here the SHIMM design and methodology for estimating key atmospheric parameters will be presented, as well as initial field test results with comparisons to the Stereo--SCIDAR instrument. 
\end{abstract}

\begin{keywords}
atmospheric effects -- site testing -- instrumentation: miscellaneous
\end{keywords}



\section{Introduction}
Atmospheric turbulence induces rapidly changing distortions and motion of stellar images from ground--based telescopes. This means for short exposure images the \ac{PSF} will become `speckled'. In the long exposure regime these average to produce the seeing limited \ac{PSF} with a \ac{FWHM} (i.e. the `seeing angle') of typically 0.5--2~arcsec at a good observing site. Turbulence at high altitudes also induces intensity fluctuations of the starlight, known as `scintillation'. The angular resolution and photometric \ac{SNR} for astronomical imaging therefore depend on the properties of the atmosphere during an observation. The total optical turbulence strength can also be characterised in terms of the optical coherence length or \ac{r0}. 

It is important to emphasise that the altitude and strength of turbulent layers affect observations in different ways. Seeing results from all turbulent layers in the atmosphere, whereas scintillation predominately results from high-- or strong mid--altitude turbulence. Therefore, it is possible to have poor seeing (i.e. a large seeing angle) but low scintillation noise if the integrated atmospheric turbulence is dominated by low altitudes. Knowledge of the vertical \ac{OTP}, as well as the overall seeing quality, is therefore useful for quality control and queue--scheduling of observations, as well as for site characterisation and selection. In most methods of \ac{AO} corrections, the \ac{OTP} also determines the \ac{theta} for effective correction. The \ac{tau} of the atmosphere, determined by the wind speed associated with the turbulent layers, is also a critical parameter for AO-assisted observations.  

The most commonly used seeing monitor is the \ac{DIMM}, which typically utilises a small auxiliary telescope at an observing site. The \ac{DIMM} measures the differential motion between images of a bright target star produced by two sub--apertures defined by a telescope aperture mask. Since it employs a differential method, measurements are insensitive to tracking errors, telescope shake, or other static optical aberrations which have an equal effect on the two images \citep{Sarazin89, Wilson99}. However, the turbulence causes small differential motions of the images \citep{Donovan03}. The differential motion is usually calculated from the centroids of the pixel intensity values for each of the two images formed on the detector. The variance of the differential motion yields an estimate for \ac{r0}. The \ac{DIMM} monitor is sensitive to bias by the effects of scintillation. Strong scintillation due to high--altitude turbulence reduces the observed differential image motion for the small sub--apertures, so that the seeing angle is systematically underestimated \citep{Tokovinin07A}. The classical \ac{DIMM} does not provide a profiling capability, therefore an external measure of the OTP would be needed to correct this effect. In addition, accurate application of the DIMM method to estimate r$_{0}$ requires careful estimation of the noise level in the centroid values resulting from the shot noise of the signal and detector noise. 

A number of variations of the \ac{DIMM} design have been developed previously. For example, the \ac{GDIMM} \citep{Aristidi14}, which employs a 3-hole aperture mask and measures seeing in the same way. The \ac{H-DIMM} \citep{Bally96} employs a Hartmann mask in order to utilise more of the telescope aperture, with a larger number of sub--apertures. 

The \ac{SHIMM} is a further development of the DIMM principle, employing a \ac{SHWFS} instead of an aperture mask with isolated sub-apertures. The lenslet array of the \ac{SHWFS} divides the re--imaged aperture of the telescope into a grid of sub--apertures.  The SHIMM therefore utilises more of the telescope aperture than the traditional DIMM, reducing the statistical noise for seeing measurements. The SHIMM provides a low--resolution estimate of the \ac{OTP}, determined from the overall seeing strength, scintillation index and the correlation of the scintillation--induced intensity fluctuations between the sub--apertures of the \ac{WFS}. This allows the effect of scintillation on the estimate of \ac{r0} to be corrected. Equipped with a suitable high frame rate detector, the SHIMM can also estimate \ac{tau}, via measurement of the power spectrum of the atmospherically--induced defocus of the wavefront at the telescope aperture. 

The SHIMM has been developed as a low--cost, compact and portable seeing monitor that can be duplicated easily and inexpensively.  It is therefore well suited for comparisons of the atmospheric conditions around a large observing site or between two or more sites. The relevant theory describing atmospheric optical turbulence and its effects on astronomical imaging will be summarised in section~\ref{section:turbulence_theory}. The design and technical details of the SHIMM instrument are described in section \ref{section:Instrument}. The methodology used to estimate key atmospheric parameters, i.e. \ac{r0}, \ac{tau}, \ac{theta}, and the results of numerical simulations are presented in section \ref{section:analysis}.  Finally, the on-sky performance of the SHIMM is discussed in section \ref{sec:field-test-results}, along with comparisons with \ac{Stereo-SCIDAR}, a high-resolution profiling instrument.

\section{Optical Turbulence Parameters}\label{section:turbulence_theory}
For optical propagation through turbulence, characterised by Kolmogorov statistics, the vertical profile of optical turbulence strength as a function of the height of the turbulent layer above the observatory $h$ (hereinafter referred to as altitude) is defined by the refractive index structure parameter $C_{n}^{2}(h)$. The total integrated turbulence between two altitudes is defined as $J$ = \(\int_{h_{1}}^{h_{2}} C_{n}^{2} \,dh\). Integrating the optical effects of the turbulence over the full extent of the atmosphere leads to the definition of the coherence length or Fried parameter \citep{Hardy98}:
\begin{equation}
\label{eq:r0_Cn}
r_{0} = \left( 0.423 k^{2} \sec(Z) \int^{\infty}_{0} C^{2}_{n} \left(h\right) \,dh \right)^{- \frac{3}{5} } \,  ,
\end{equation} 
where $k = 2\pi/\lambda$ is the wavenumber, $\lambda$ is the wavelength and $Z$ is the zenith angle. 
The Fried parameter is a key measurement for characterising atmospheric seeing. It defines the maximum telescope pupil diameter for which the angular resolution remains diffraction limited in the presence of optical turbulence \citep{FohringThesis}. For all telescopes with diameters larger than $r_{0}$ the angular resolution (FWHM of the long exposure point spread function) is seeing limited and is given by:
\begin{equation}
\Omega_{\mathrm{FWHM}} = 0.98\frac{\lambda}{r_{0}} \, .
\label{eq:seeing_lim}
\end{equation} 
\noindent The isoplanatic angle is the angular size of the sky over which the seeing--induced optical aberrations may be considered approximately uniform and is defined as 
\begin{equation}
\label{eq:theta}
\theta_{0} = 0.314 \frac{r_{0}}{h_{\mathrm{eff}}} \, ,
\end{equation} 
where $h_{\mathrm{eff}}$ is the effective turbulence altitude, defined as
\begin{equation}
\label{eq:heff}
h_{\mathrm{eff}} = \left[\frac{\int^{\infty}_{0} C^{2}_{n} \left(h\right)h^{\frac{5}{3}} \,dh} {\int^{\infty}_{0} C^{2}_{n} \left(h\right) \,dh} \right]^{\frac{3}{5}}\, .
\end{equation} 
The coherence time is a measure of the timescale of the optical aberrations due to turbulence, defined as 
\begin{equation}
\label{eq:tau0}
\tau_{0} = 0.314 \frac{r_{0}}{v_{\mathrm{eff}}} \,  ,
\end{equation} 
where $v_{\mathrm{eff}}$ is the effective wind velocity of the turbulence, defined as  
\begin{equation}
\label{eq:veff}
v_{\mathrm{eff}} = \left[\frac{\int^{\infty}_{0} C^{2}_{n} \left(h\right)V\left(h\right)^{\frac{5}{3}} \,dh} {\int^{\infty}_{0} C^{2}_{n} \left(h\right) \,dh} \right]^{\frac{3}{5}} \, ,
\end{equation} 
where $V(h)$ denotes the velocity profile with altitude. The isoplanatic angle and the coherence time are key parameters for the application of \ac{AO} correction for astronomy. The isoplanatic angle defines the field of view over which wavefront corrections, determined for a single reference direction, will be valid. The coherence time determines the required minimum temporal sampling of the \ac{AO} for effective correction.    

Scintillation is the spatio--temporal intensity fluctuation that results from the optical propagation of wavefronts which have acquired phase aberrations due to atmospheric turbulence. Propagation of the aberrated wave creates a pattern of spatial intensity fluctuations, known as `flying shadows', across the telescope aperture. With respect to astronomical photometry, this creates random fluctuations of the measured intensity as light is deviated into or out of the telescope aperture. Since wind moves the turbulence across the field of view of the telescope, this causes temporal fluctuations of the total integrated intensity of the image \citep{Osborn11}. 

The magnitude and spatial scale of the intensity fluctuations due to scintillation increase with the strength and propagation distance to the telescope, and hence the altitude of the turbulent layers. The interference of the wavefront with itself creates a pattern of `flying shadows' in the pupil plane, the characteristic scale of which is determined by the Fresnel radius $(r_{F}=\sqrt{\lambda h\sec(Z)})$. As the altitude increases so does the spatial scale of these patterns, as well as the intensity variations at the pupil plane. The magnitude of the optical intensity fluctuations due to scintillation is quantified by the normalised variance of the signal, or scintillation index 
\begin{equation}
\label{eq:scint_index_threory}
\sigma ^{2}_{I} = \sum \frac{ \langle  I^{2}  \rangle - \langle  I  \rangle ^{2} }{ \left \langle I\right \rangle ^{2}} \, ,
\end{equation}
where $I$ is the intensity and $\langle \rangle$ denotes time averaging. The RMS photometric noise (fractional intensity fluctuation) due to scintillation is given by $\sqrt{\sigma ^{2}_{I}}$ \citep{Osborn15}.

Scintillation noise makes a significant contribution to the overall uncertainty of photometric measurements with ground--based telescopes \citep{FohringThesis}. The scintillation noise variance is greatly reduced by spatial averaging over a large telescope aperture and by temporal averaging over a long exposure. However, since shot noise is also reduced similarly, scintillation remains the limiting factor in the precision of photometric measurements of bright stars, in all cases. Hence the scintillation index is also a key parameter for atmospheric characterisation for astronomy.  

\section{The SHIMM Instrument} \label{section:Instrument}

\begin{figure} 
\begin{center}
\includegraphics[width=.6\columnwidth]{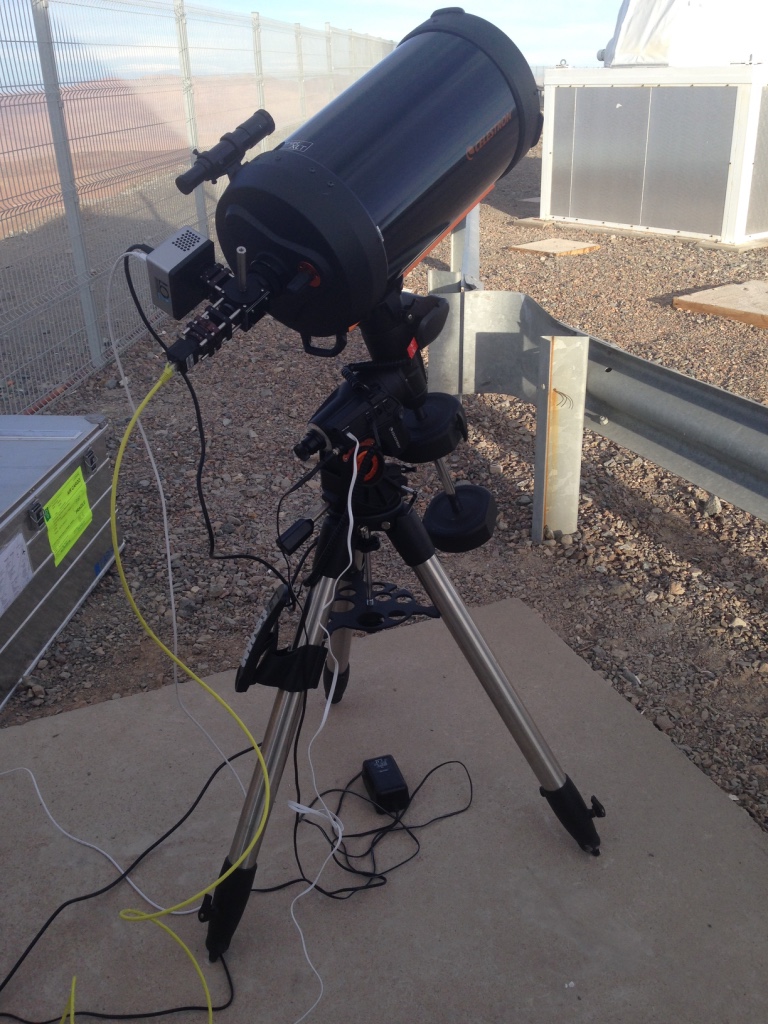}
\end{center}
\caption{The SHIMM (with additional FASS optics) at Cerro Paranal, Chile, at the site of the VLT \citep{Guesalaga16}.} \label{fig:shimm_paranal_photo}
\end{figure}

The \ac{SHIMM} instrument comprises a telescope equipped with a \ac{SHWFS} module. Figure \ref{fig:shimm_paranal_photo} shows an image of the \ac{SHIMM} (with additional \ac{FASS} optics) at Paranal Observatory, Chile. \ac{WFS} images are recorded for bright star targets using very short exposures ($\sim$1--2~ms) at a frame rate of a few tens of Hz to sample the statistics of the rapidly changing atmospheric optical aberrations. An important goal was to develop an instrument that is easily portable, at a relatively low cost and that could be easily replicated. Hence the SHIMM is based on small aperture telescopes and exploits off--the--shelf components. Table \ref{table:SHIMM_instruments} summarises the components and hardware parameters of the prototype SHIMM instrument.

\begin{table}
    \centering
    \begin{tabular}{p{2.5cm}|p{5.cm}}
	\hline
	\textbf{Features} & \textbf{Specifications}  \\
    \hline
    \multicolumn{2}{|p{7cm}|}{\textbf{TELESCOPE}}\\
	\hline  
	Model: & Celestron CGEM 9.25 inch f/10 Schmidt Cassegrain\\
	Aperture: & 9.25 inches \\
    f number: & f/10 \\
	Mount: & VX Mount \\
    Mount Operation: & Durham SLODAR scripts operated on the SHIMM NUC computer \\ 
	Autoguiding & RS232 controlled relay board implementing offsets provided by WFS measurement\\    
	\hline

    \multicolumn{2}{|p{7cm}|}{\textbf{SHIMM}}\\
	\hline
	\textbf{Optics} & \\
	Collimator: & {Achromat lens with focal length of 30 mm}\\
	Lenslet Array: &{Lenslet array with pitch of 0.5 mm and focal length of 15.3 mm}\\

	Lens Cage:& {Lens mounts and translation and rotation stages}\\
	\hline
	\textbf{Detector} & \\
	Model:  & 1288 x 728 Mono Point Grey 092SM-CS Blackfly GigE camera\\
	Lens Mount: & CS-Mount \\
	Pixel Size: & 4.08 $\mu$m \\
	Frame Rate: & 30 Hz\\
	Read Noise: & 8.28 e$^{-}$ \\
	Quantum Efficiency: & 52 $\%$ (at 525 nm)\\
 
	ROI: &  728 $\times$ 728 \\
	Exposure Time: &  2 ms \\	
	Image Scale: & 0.71 arcsec/pixel\\  	
	\hline
	\textbf{PC} & \\
	Hardware: & Mini PC Intel Nuc\\
	OS: & Ubuntu 12.0 \\
	\hline
	\textbf{sub--apertures} & \\
	Size: &  4.1 cm \\
	$\#$ used:  & 12 \\
	\hline

    \end{tabular}
    \caption{Hardware specifications for the prototype SHIMM.}
    \label{table:SHIMM_instruments}
\end{table}

\begin{figure} 
\begin{center} 
\includegraphics[width=1.\columnwidth]{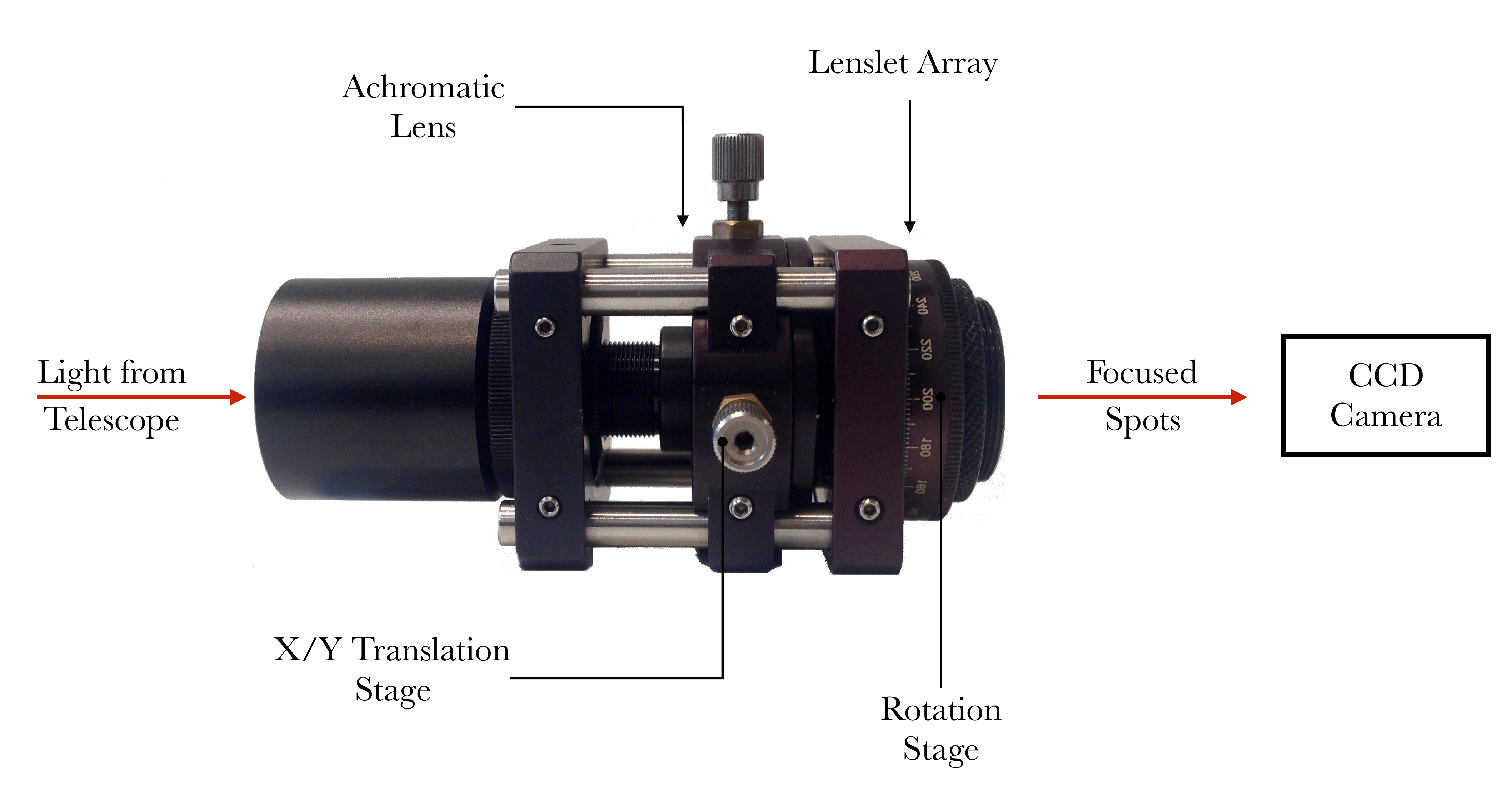}
\end{center}
\caption{Image of the SHIMM WFS optics, comprised of an achromatic lens and lenslet array. The orientation of the spot pattern imaged onto the CCD can be altered with the rotation mount. The alignment of the  lenslet array with respect to the telescope aperture is adjusted using the translation stage.}\label{fig:shimm_optics}
\end{figure}

The configuration of the \ac{SHWFS} is shown in figure \ref{fig:shimm_optics}. Light gathered from the target star at the focus of the telescope is collimated by an achromatic lens. A lenslet placed at the optical conjugate of the telescope primary mirror divides the projected telescope aperture area into a grid of sub-apertures and focuses the resulting array of sub--images onto the detector. A small translation stage is used to centre the lenslet array relative to the projection of the telescope aperture. A rotation stage then allows the \ac{WFS} image pattern to be aligned with respect to the detector. 

A key aspect of the SHIMM design is to define and optimise the pattern of \ac{WFS} sub--apertures projected across the telescope aperture. The \ac{WFS} must provide sufficient sampling, in terms of the number of sub--apertures across the telescope aperture, to measure the lowest order Zernike modes of the turbulent aberration, including the second order defocus mode. However, increasing the number of \ac{WFS} sub--apertures reduces their individual projected diameter for a given telescope aperture size. For smaller sub--apertures the effective area and hence the signal acquired in each sub--image is reduced, and the angular size of the \ac{WFS} sub--images (or `spots') due to diffraction increases. This results in poor \ac{SNR} of the centroid measurements, and small image motions relative to the \ac{FWHM} of the spots themselves. In addition, it is desirable to maximize the fraction of the telescope aperture area utilised for \ac{WFS} measurements, and with minimal vignetting of individual \ac{WFS} sub--apertures by the edges of the aperture and by secondary mirror obscuration. 

\begin{figure}
\begin{center} 
\includegraphics[width=1.\columnwidth]{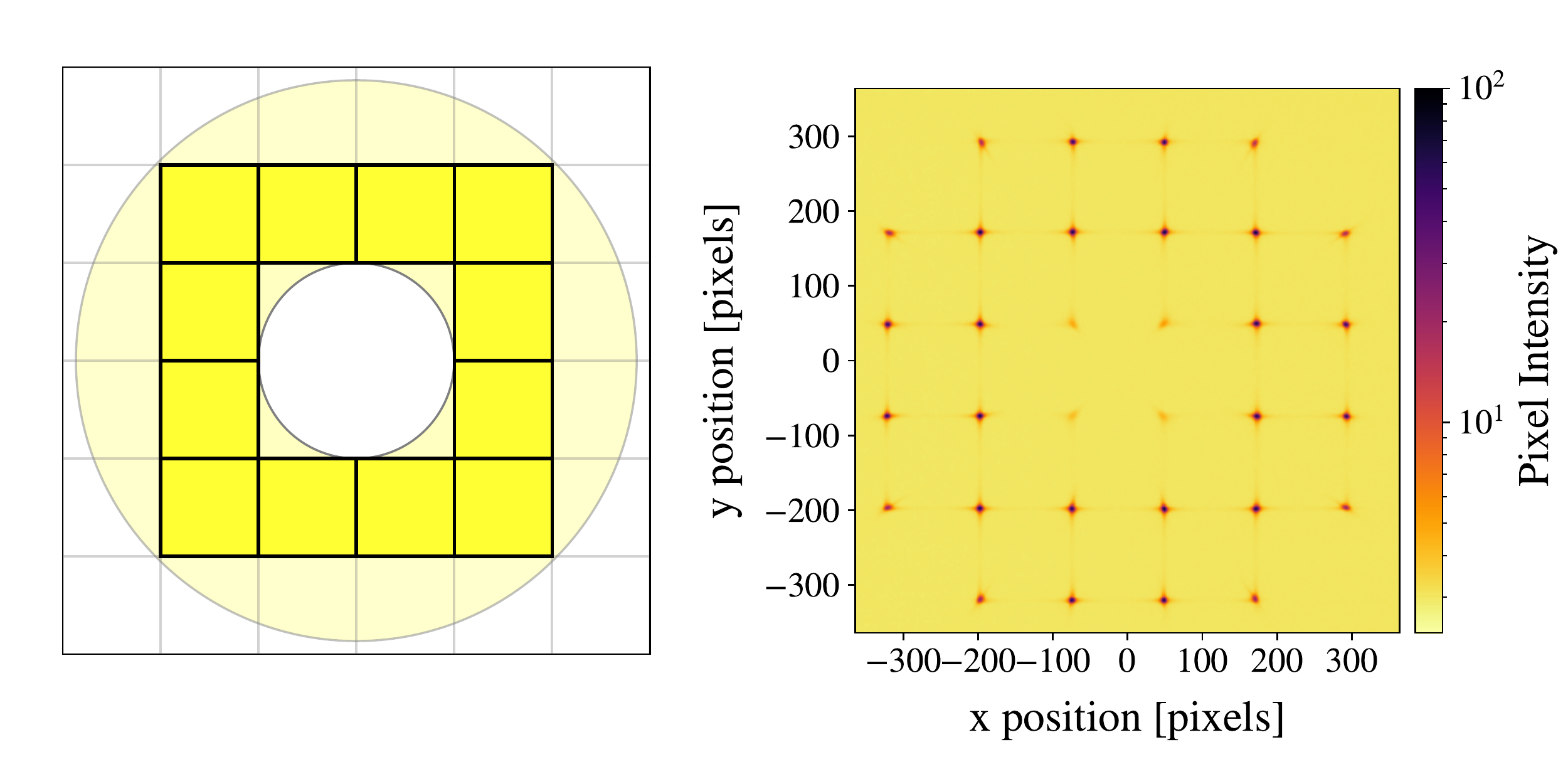}  
\end{center}
\caption{(Left)  Illumination pattern of the square sub--apertures of the WFS mapped onto the SHIMM telescope aperture. The outer and inner circles indicate the edges of the primary and secondary mirrors. The dark line grid indicates the fully illuminated sub--apertures. (Right) The resulting spot pattern formed on the detector. Note, the image has been stretched to highlight the pattern. \label{fig:subapertures}}
\end{figure}

The \ac{WFS} projection chosen for the prototype SHIMM instrument is shown in the left panel figure \ref{fig:subapertures}. For this analysis, only the fully--illuminated central 12 sub--apertures were used. However, depending on the degree of vignetting, additional sub--apertures could be included.  The right panel of figure \ref{fig:subapertures} shows the resulting \ac{WFS} image for a bright target star. When deployed on the 9.25~inch telescope, for this \ac{WFS} configuration each sub--aperture has a projected length of 4.1~cm. There is then an adequate \ac{SNR} for wavefront sensing for target stars of magnitude V~=~2 or brighter. For this limiting magnitude, and elevations above 60$^{\circ}$, it was calculated that $96 \%$ and $83 \%$ nighttime sky coverage is possible at Roque de los Muchachos Observatory, La Palma, and Paranal Observatory, Chile, respectively. Lower elevation results in increased turbulence strength and scintillation. However, by reducing the minimum elevation or using slightly fainter targets continuous nighttime observations are possible.

The \ac{SHIMM} requires short exposure images at a frame rate that can sample the changing atmosphere. It therefore needs a detector with a relatively low readout noise and a frame rate at least of the order of a few tens of Hz. With these criteria and a low-cost requirement, the 092SM-CS Blackfly camera was chosen for the prototype SHIMM. 

Software for data acquisition, real--time data analysis and the display was developed for the SHIMM and operated on a compact ITX mini PC running a 
Linux operating system. Target acquisition was performed manually but auto--guiding of the telescope was implemented automatically using offsets calculated from the global position of the WFS pattern on the detector. 

Wind--shake and vibration can be problematic for seeing instruments using small portable telescopes, including the \ac{DIMM} and \ac{SHIMM}. Since a differential image motion method is used, the measurements are not biased by small telescope guiding and wind shake errors. However, high wind speeds can result in very fast motions, so that the \ac{WFS} images are significantly `smeared' within a single exposure. Large excursions may also cause the \ac{WFS} pattern to be lost from the field of view entirely. To reduce these effects a portable windbreak enclosure was employed to shield the SHIMM monitor during high local winds.

\section{Data Analysis}\label{section:analysis}

\subsection{Estimating The Fried Parameter ($r_{0}$)} \label{section:estimater0}

The Fried parameter can be estimated from the SHIMM \ac{WFS} data using the method described by \citet{Butterley06} for the \ac{SLODAR} instrument, where a theoretical model is fitted to the time--averaged auto--covariance of the centroid values for a range of spatial separations within the WFS array. The theoretical auto--covariance map can be generated via numerical integration for a given \ac{WFS} geometry and sub--aperture size. 

For small sub--apertures, such as those used in the prototype SHIMM instrument, the shape of the auto--covariance function will be affected by scintillation such that a correction is required, as described in section \ref{sec:corrr0}. In the first instance, the measurement of $r_{0}$ without taking into account the effects of scintillation on the auto--covariance of the centroids is described. 

SHWFS images are typically recorded in packets of a few hundred frames at a frame rate of a few tens of Hz, with an exposure time of 1--2~ms. Each data set, which yields a single seeing measurement, comprises one or more sequential packets spanning a duration of a few tens of seconds. Due to the large number of frames used for a single measurement ($N \sim 300 - 1500$), the statistical noise will be $\sim 1/\sqrt{N} \sim 3-5 \%$. The centroids are determined for each WFS spot within each frame of the dataset by using the standard centre-of-mass equation. A sub-region of pixels is defined for each spot and an intensity threshold is applied (such that everything below is set to zero) to reduce the influence of readout noise on the centroid measurements. Applying an intensity threshold can result in an inaccurate estimate of \ac{r0}. Therefore, the choice of the threshold was optimised for the prototype SHIMM to minimise this effect. This resulted in a less than 1 $\%$ overestimate of \ac{r0} for values greater than 0.1 m.

Common spot motions due to wind shake and telescope guiding errors are removed by subtracting the mean of the centroids over all WFS spots for each frame. As a result, common tip-tilt motions induced by the atmosphere are also removed. The calculation of the theoretical auto--covariance map also assumes that common motions are fully removed \citep{Butterley06}. The mean centroid for each sub--aperture over the duration of the data set is then subtracted, in order to remove any static or very slowly varying aberrations of the telescope and WFS optics. Under typical wind speeds, the centroids for small WFS sub--apertures are expected to average to zero within seconds.

\begin{figure}
	\includegraphics[width=1.\columnwidth]{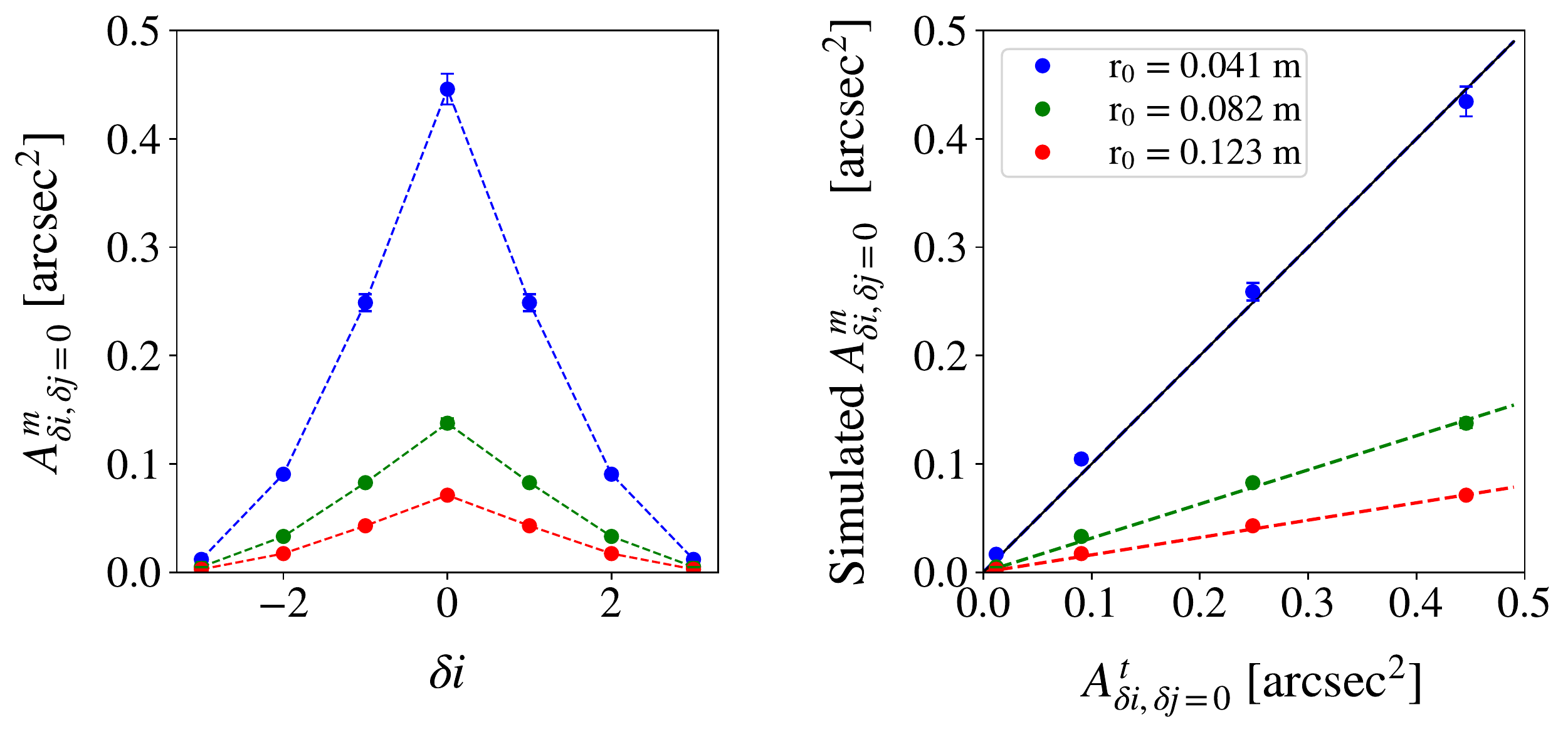} 
    \caption{Auto-covariance values for simulated data, assuming Kolmogorov turbulence, for centroids in one dimension only. (Left) A 1-D slice of the auto-covariance map at $\delta j$ = 0. (Right) The relationship between the theoretical auto-covariance ($A^{t}_{\delta i, \delta j}$) and the auto-covariance for simulated WFS data ($A^{m}_{\delta i, \delta j}$), where $\delta j$ = 0 and $\delta i$ = 0,1,2,3. The blue, green and red markers indicate results for different $r_{0}$ values of 0.041 m, 0.082 m and 0.123 m respectively. The broken lines indicate the linear fit and the black solid line is x=y. The error bars represent the statistical uncertainty of the simulation, which is the only noise source.  \label{fig:get_r0}}
\end{figure}

Finally, the spatial auto-covariance of the x-- and y--centroids are calculated separately as
\begin{equation}
A_{\delta i, \delta j} = \langle C_{i,j} C'_{i',j'} \rangle \, ,
\end{equation}
where $C$ and $C'$ are the centroids at sub--aperture position $[i,j]$ and $[i',j']$ respectively. The spatial offsets between the sub--apertures, in units of the sub--aperture diameter, are given as $\delta i$ and $\delta j$. The auto--covariance map is created by calculating this for every possible sub--aperture separation of the WFS array. 

The value of \ac{r0} is determined by fitting the theoretical auto--covariance model to a one--dimensional slice cut through the covariance map at $\delta i$ = 0 or $\delta j$ = 0 for x-- and y--centroids respectively.  
In principle, the fit can be made to the full auto--covariance map in two dimensions. However, the \ac{SNR} of the covariance map (which is dominated by statistical noise) reduces as you move away from the centre. From numerical simulations it was found that utilising the full auto--covariance did not improve the precision of seeing measurements with the SHIMM for data packets of a few seconds duration. Figure \ref{fig:get_r0} shows examples of the one--dimensional covariance maps for simulated SHIMM data, demonstrating the expected variation as $r_{0}^{-5/3}$.  

The fit to the auto--covariance can be linearized by plotting the measured ($A^{m}_{\delta i, \delta j}$)  versus theoretical ($A^{t}_{\delta i, \delta j}$) auto--covariances. Noting that the auto--covariance is symmetrical about $\delta i = 0$, then 
\begin{equation}\label{eq:fit}
A^{m}_{\delta i, \delta j}  =  A^{t}_{\delta i, \delta j}  \left(\frac{r_{0}}{d}\right)^{-\frac{5}{3}} \, ,
\end{equation} 
where $d$ is the size of the sub--aperture. The linear relationship is shown in figure \ref{fig:get_r0} for simulated SHIMM data. 

\begin{figure}
\includegraphics[width=1.\columnwidth]{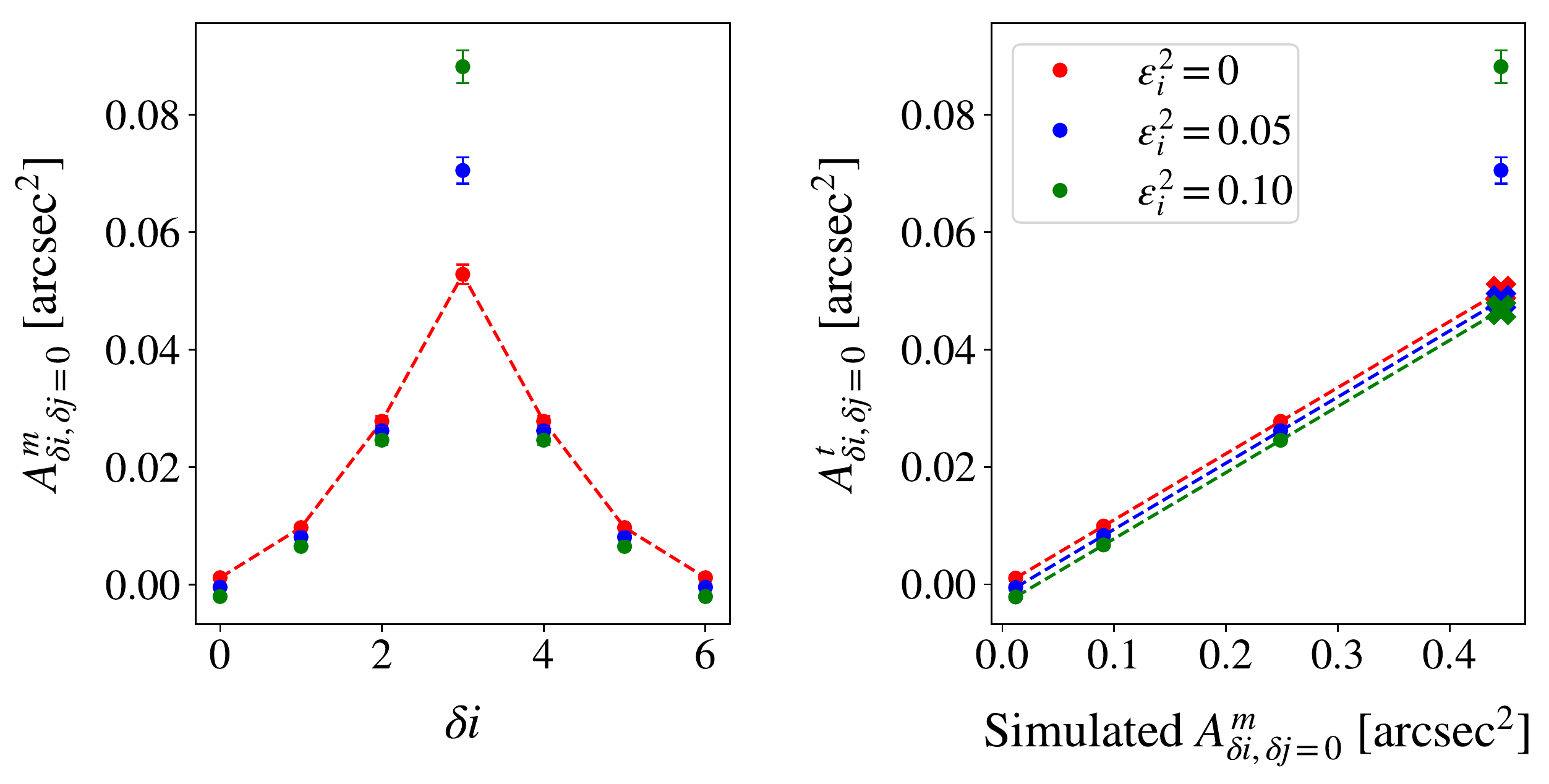} 
\caption{(Left) Example $A^{m}_{\delta i, \delta j}$, where $\delta j$ = 0, for simulated WFS data including shot noise. The dashed red line indicates the fit to the data in the case of zero noise. (Right) Shows the effect noise has on the linear fit between $A^{t}_{\delta i, \delta j}$ and $A^{m}_{\delta i, \delta j}$, where $\delta j$ = 0. The dashed lines indicate the linear fit when the measured zero-offset point ($j = i$) is excluded from the fit. The crosses indicate the estimated centroid variance in the absence of shot noise. \label{fig:get_r0_noise}}
\end{figure}

Significant sources of noise in measuring the centroids of the WFS spots include the shot noise of the signal and detector readout noise. Since short exposure images are used with bright sources, the effects of sky background light and dark current are negligible. Any error introduced by removing the individual mean centroid will also contribute to the noise. The noise contribution to the auto-covariance is given by
\begin{equation}
   \langle \epsilon_{l}\epsilon_{k} \rangle =
  \begin{cases}
    \left( 1 - \frac{1}{n} \right)\langle \epsilon_{l}^{2} \rangle & \text{if $l = k$} \\
     - \frac{1}{n} \langle \epsilon_{l}^{2} \rangle & \text{if $l \neq k$}
  \end{cases} \label{eq:noise} \, , 
\end{equation}
where $\epsilon$ is the noise in each centroid and subscripts $l$ and $k$ refers to the sub--aperture positions of $[i,j]$ and $[i',j']$. These describe the slopes for one axis (i.e. x-- or y--centroids). From equation \ref{eq:noise}, the effect of noise decreases as the number of sub--apertures of the WFS is increased, except in case of $l = k$ (i.e. the centroid variance), as illustrated in figure \ref{fig:get_r0_noise}. For this reason, the central point 
from the auto--covariance fit is excluded so that the effect of noise on the estimate of \ac{r0} is greatly reduced. The centroid noise level can be estimated and monitored (for data quality control) by taking the difference between the measured centroid variance and its expected value extrapolated from theoretical fit to the auto--covariance. Since the prototype SHIMM only uses 12 sub--apertures,  according to equation \ref{eq:noise}, for auto--covariances where $\delta i \neq \delta j$ the value of the auto--covariance will be reduced by $ - \frac{1}{n} \langle \epsilon_{l}^{2} \rangle$. Figure \ref{fig:get_r0_noise} demonstrates this for different noise levels. This means that the fit will no longer intercept at the origin but will be offset by $ - \frac{1}{n} \langle \epsilon_{l}^{2} \rangle$. However, the gradient of the fit will remain unchanged, as demonstrated in the left subplot of figure \ref{fig:get_r0_noise}.

\subsection{Estimating the Optical Turbulence Profile (OTP)} \label{section:Estimating_Profile}

The \ac{SHIMM} provides a low--resolution estimate of the \ac{OTP} by measuring the scintillation index for the WFS spots, together with the correlation of the scintillation intensity fluctuations between neighbouring sub--apertures. The use of scintillation intensity correlations in estimating the \ac{OTP} from \ac{SHWFS} data has previously been exploited by the optical turbulence profiler \ac{CO-SLIDAR} deployed on a large ($\sim$1.5 m) telescope. \ac{CO-SLIDAR} is a crossed-beams method, similar to SLODAR, in which the \ac{OTP} is recovered from the cross--covariance of both the centroids and the intensities of the WFS spots for a double star target \citep{Robert11}. Another development, known as \ac{SCO-SLIDAR}, exploits the auto--covariances of both the centroids and intensities for a bright, single star target, to estimate the \ac{OTP} using a WFS applied to a small telescope, similar to the SHIMM \citep{Vedrenne07}. The main difference to the SHIMM is that \ac{SCO-SLIDAR} uses smaller sub--apertures so that the scintillation signal is stronger and is correlated over larger WFS separations, but a very bright target is needed and the \ac{FWHM} of the WFS spots are larger and therefore will be less sensitive to the image motions due to turbulence. For the SHIMM larger sub-apertures are used, resulting in weaker scintillation. However, the larger sub--apertures permit the use of fainter targets and hence continuous monitoring. 

Atmospheric turbulence comprises contributions from layers at a range of altitudes, often including a strong ground layer. High--altitude turbulence results in the overestimation of \ac{r0}. Therefore, it is not possible to distinguish whether a measured value of \ac{r0} results from a relatively strong but high altitude layer or a weaker low altitude layer. For example, for the prototype SHIMM, a measured \ac{r0} of 0.1~m could be due to a turbulent layer at the ground with \ac{r0} = 0.1~m or a turbulent layer at \mbox{16~km} altitude with \mbox{\ac{r0} = 0.065~m}.

For \ac{SHIMM} data, measuring $\sigma_{I}^{2}$ and Corr makes it possible to distinguish between the scenarios described above. Both the correlation and the scintillation index increase with propagation distance. However, the correlation additionally increases with decreasing \ac{r0} values. The scintillation index is calculated using equation \ref{eq:scint_index_threory}. The correlation between neighbouring sub--apertures is given by 
\begin{equation}
\mathrm{Corr} =  \frac{\sum (A \times B) }{\sqrt{\sum A^{2} \times \sum B^{2}}}  \, ,
\end{equation}
where  
\begin{align}
A =  I_{A} - \langle I_{A} \rangle \, , \\
B =  I_{B} - \langle I_{B} \rangle \, ,
\end{align}
and $I$ is the time-varying intensity of a single sub--aperture $A$ or $B$. The Corr values for the SHIMM data are calculated by averaging the correlation over all instances where $A$ and $B$ are neighbouring sub--apertures in either the x-- or y--direction. 

A three--layer model at chosen fixed altitudes was created, empirically through simulation, such that three unknowns (the turbulence strength at these altitudes) are estimated from three measurable quantities; $\sigma_{I}^{2}$, Corr and the pre-corrected estimate of \ac{r0}. Here the model is defined with layer altitudes of 0, 5 and 15 km, chosen so each layer produces a distinguishable response in the SHIMM measurements i.e. they have a close to orthogonal response in the fit. In more detail:
\begin{itemize}
\item \textbf{The Ground Layer (0 km):} The ground layer turbulence does not cause any scintillation effects, but does contribute to the total integrated seeing and hence the magnitude of the centroid covariance. Furthermore, typically there is always significant optical turbulence at the surface level, resulting from the interaction of the wind with the ground, and the heating or cooling effect of the ground on the air above it.  
\item \textbf{The Higher Layer (15 km):} Higher altitude layers produce strong scintillation and spatial intensity fluctuations on relatively large scales (a few cm) so there will be a larger value of $\sigma_{I}^{2}$ as well as a significant correlation of the intensities between neighbouring sub--apertures of the SHIMM. Typically, there is also strong turbulence at the altitude of the jet stream, in the region between 10~km and 20~km. 
\item \textbf{The Middle Layer (5 km):} 
Turbulence at intermediate altitudes, approx. 3~km to 8~km, will produce moderate intensity fluctuations due to scintillation but without significant correlation between neighbouring sub--apertures of the SHIMM. Hence the SHIMM measurements support the inclusion of a third, intermediate, layer in the model, which is placed at 5~km. 
\end{itemize}

\begin{figure}
\begin{center}
\includegraphics[width=.8\columnwidth]{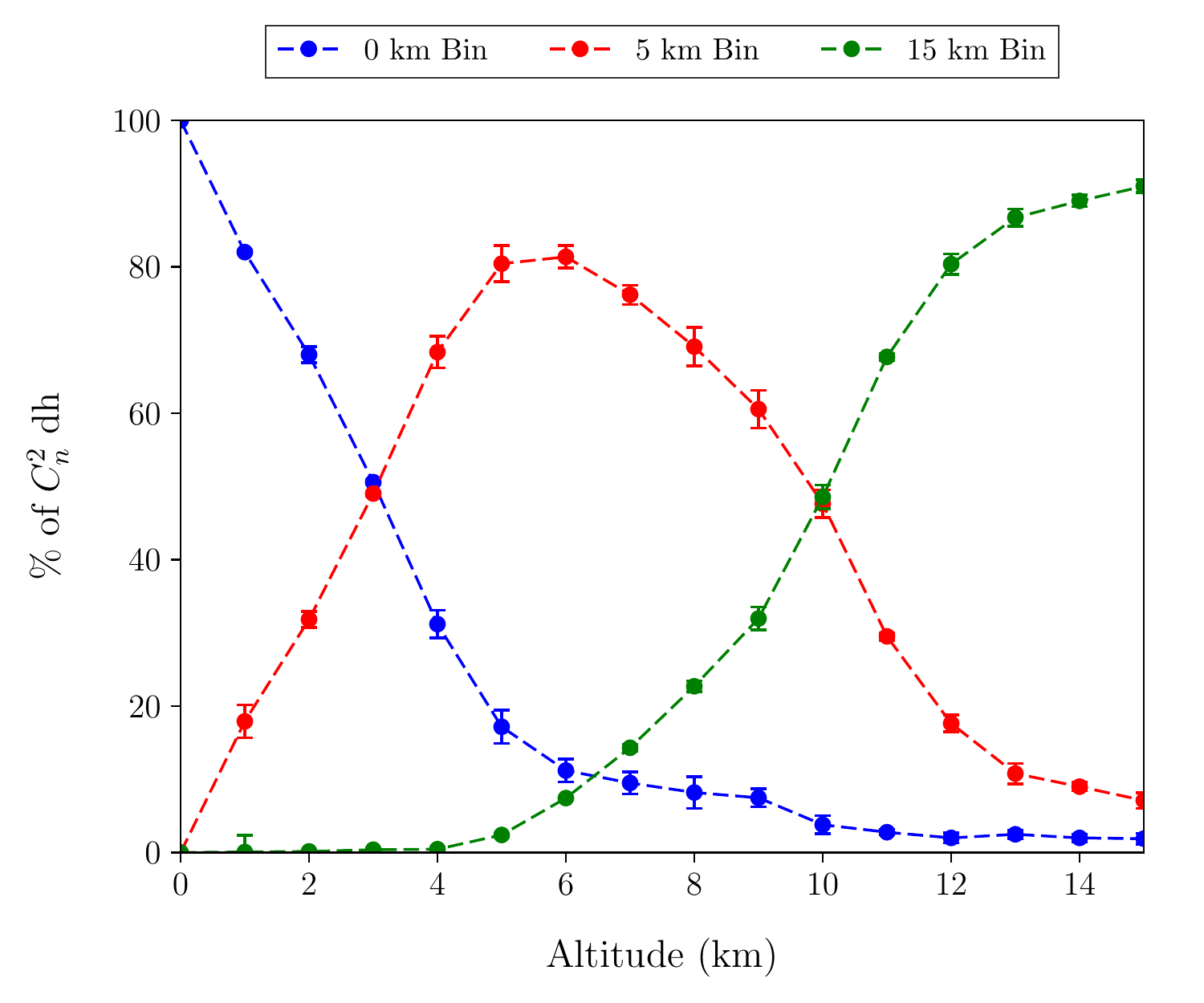}
\end{center}
\caption{Illustration of the response of the SHIMM three--layer model of the OTP for a  single turbulent layer at a range of altitudes, from numerical simulation of the prototype SHIMM instrument. As a single layer is placed at different altitudes the total turbulence strength of that layer is distributed amongst the three defined altitude bins, as shown.   }\label{fig:layer_placement}
\end{figure}

Figure \ref{fig:layer_placement} illustrates the response of the SHIMM three--layer model of the OTP for a single turbulent layer at a range of altitudes, for simulated data. The response is not perfectly orthogonal. For example, for a layer placed exactly at 5~km the SHIMM model places 80$\%$ of the turbulence strength in the 5~km output layer, with the remainder split evenly between the 0~km and 15~km bins. However, the model is able to distinguish between turbulence at the ground, at mid and at high altitudes sufficiently well for a number of applications, for example, to give an accurate estimate of \ac{theta} for \ac{AO}, and to correct the measurement of seeing for the effects of scintillation. In addition, a low--resolution profile estimate of the OTP would be useful for site characterisation and queue scheduling and to estimate the level of photometric noise from scintillation for telescope observations.

\subsection{Correction of the Fried Parameter ($r_{0}$) and Estimating the Isoplanatic Angle ($\theta_{0}$)}\label{sec:corrr0}

Scintillation of the light from the target star leads to a reduction in the variance of the measured centroid motions, as well as a change in the auto--covariance shape, for the small sub--apertures of the \ac{SHWFS}. This results in an overestimation of the value of \ac{r0}. The same effect is also relevant to the DIMM, which will also overestimate $r_{0}$ in the presence of scintillation. This effect is discussed for the \ac{SLODAR} instrument in \citet{Goodwin07}. 

\begin{figure}
\begin{center}
\begin{tabular}{c} 
\includegraphics[width=1.\columnwidth]{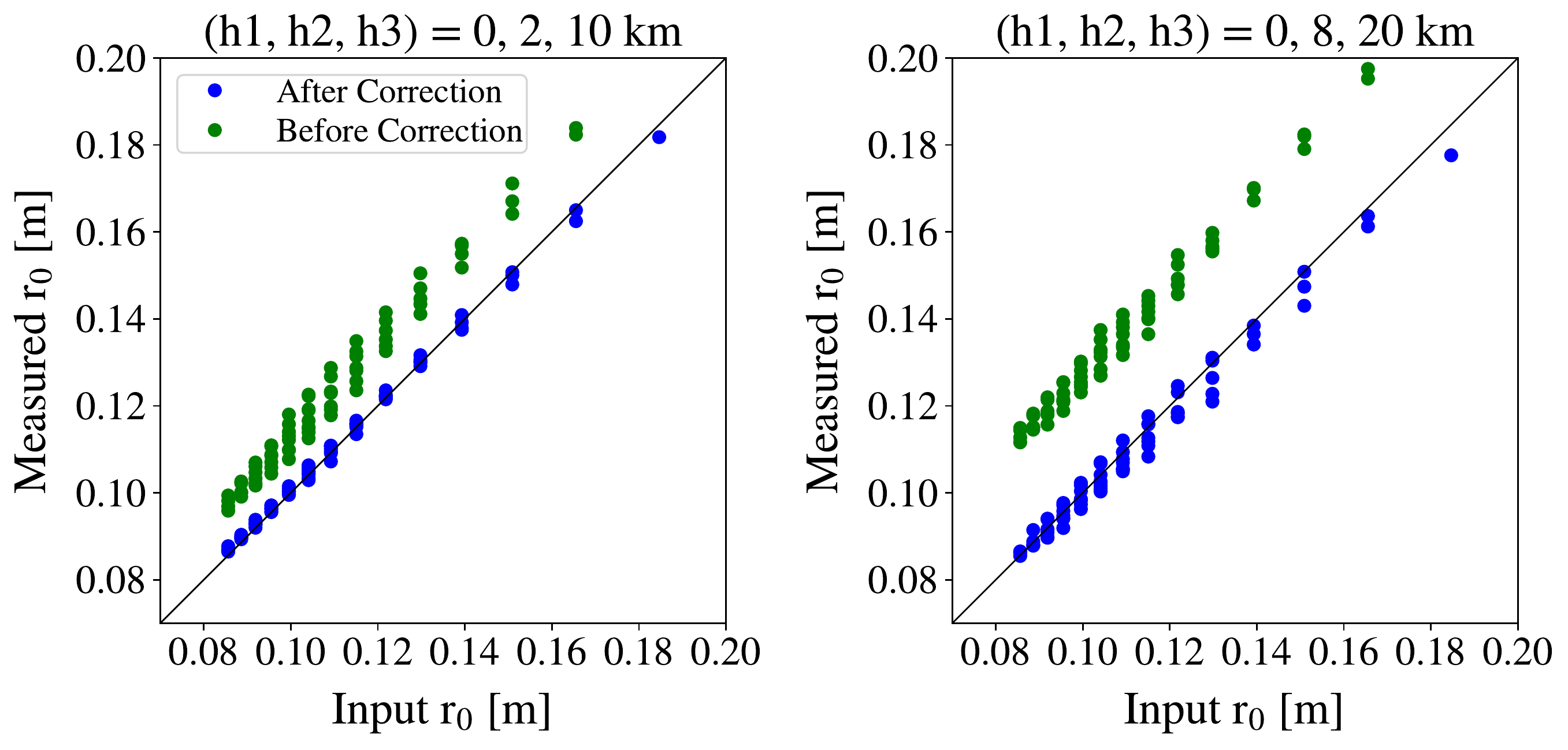}
\end{tabular}
\end{center}
\caption{Results from numerical simulations of the correction of $r_{0}$ measurements for the effects of scintillation, with  uncorrected (green) and corrected (blue) values of $r_{0}$ versus the known input $r_{0}$. Each figure displays results for two turbulent layers at varying heights and a fixed ground layer.}\label{fig:profile_3correction}
\end{figure}

The total measured integrated turbulence strength of the atmosphere, $J_{m}$, can be derived from \ac{r0}, as according to equation \ref{eq:r0_Cn}. By using the method described in the previous section, the turbulence strengths at 5 and 15 km  ($J_{5\,\mathrm{km}}$ and $J_{15\, \mathrm{km}}$ respectively) can be estimated. If the turbulence strength is known at these individual layers, the change in the measured turbulence strength of that layer induced by its propagation distance ($\Delta J_{5 \,\mathrm{km}}$ and $\Delta J_{15\, \mathrm{km}}$) can be estimated empirically, through simulation. The sum of $\Delta J_{5 \,\mathrm{km}}$ and $\Delta J_{15\, \mathrm{km}}$ is equal to the total change of the measure turbulence strength and can be subtracted from $J_{m}$ to derive the true total turbulence and therefore the corrected \ac{r0}. Figure \ref{fig:profile_3correction} shows the results of numerical simulations for the value of \ac{r0} estimated before and after correcting for scintillation, for different turbulence profiles. The plots show results for example three--layer turbulent profiles, where the altitude profile is denoted on each plot. A range of integrated turbulent strengths were used ($J$ = $100 \times 10^{-15}$ to 500 $\times 10^{-15}$ m$^{1/3}$) for turbulent layers at altitudes of \mbox{5 and 15 km} with a fixed turbulent strength of $J$ = 50 $\times 10^{-15}$ m$^{1/3}$ at altitude \mbox{0 km}. It should be noted that these profiles are intentionally high--altitude heavy for the purpose of amplifying this effect, this is not a typical profile. 

By comparing the values for \ac{r0} before and after correcting for scintillation, it can be seen that the estimated value of \ac{r0} after correction is much closer to the input \ac{r0} value. Whilst the correction is not always exact, it is at least an order of magnitude improvement in accuracy to the original estimation of \ac{r0}.

\begin{figure}
\begin{center}
\begin{tabular}{c} 
\includegraphics[width=1.\columnwidth]{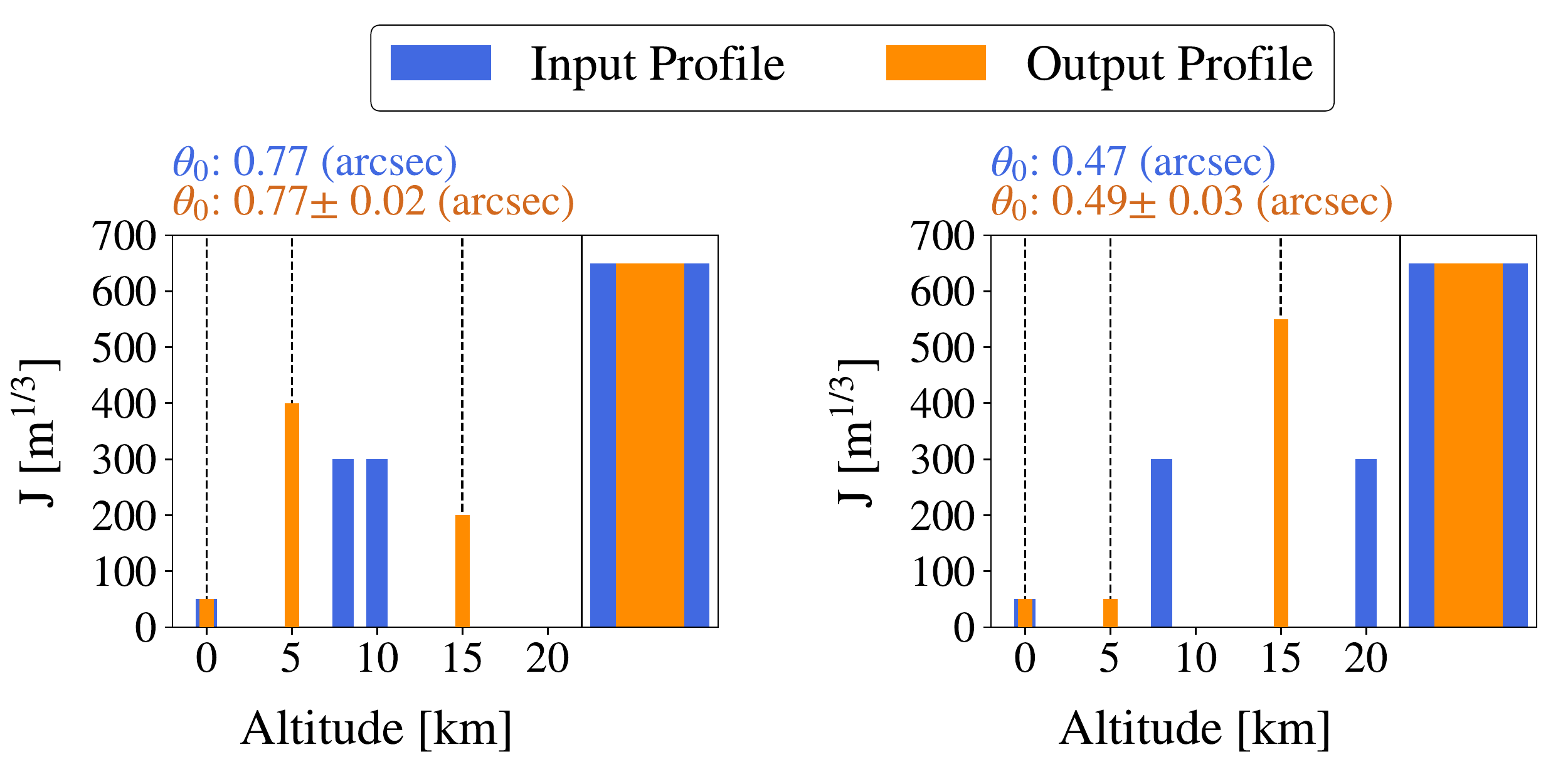}
\end{tabular}
\end{center}
\caption{Example results from numerical simulations of SHIMM measurements, for two different examples of the input turbulence profile (blue).  
The three--layer output turbulence profile (orange), with layers at altitudes \mbox{0, 5 and 15 km} (denoted by the black dashed lines) is fitted to the measured parameters of $r_{0}$, scintillation index and correlation produced by the simulation. The single broad bar on the right hand side of each figure displays the total turbulence strength. The values of $\theta_{0}$, calculated from the two profiles, are noted on the left above each figure.}\label{fig:profile_3prof_good}
\end{figure}

After correcting \ac{r0} the turbulence strength at 0 km can be derived by subtracting the strength acquired from the altitudes at \mbox{5 and 15 km} from the corrected total integrated strength. It is therefore possible to estimate a low-resolution three--layer turbulence profile and therefore \ac{theta} as according to equation \ref{eq:theta}. Figure \ref{fig:profile_3prof_good} shows simulated examples of how the input turbulence profiles are distributed in the three--layer profile. Though this is a low--resolution profile, the model can still accurately estimate the input \ac{theta} within error. The error is given by the propagation of error from the uncertainty in the fit of the theoretical model.

\subsection{Estimating Coherence Time ($\tau_{0}$)} \label{section:tau0}
Knowledge of \ac{tau} is important for characterising the atmospheric conditions since it measures how fast the turbulence is evolving. As described by equation \ref{eq:tau0}, \ac{tau} is inversely proportional to the \ac{veff}, which is defined by equation~\ref{eq:veff}. In order to estimate \ac{tau} the \ac{SHIMM} required a camera with a faster frame rate. The prototype \ac{SHIMM} was therefore upgraded to an 11~inch aperture telescope and utilised a  640 × 480 Mono Prosillica GE 680 camera. This increased the number of sub--apertures used from 12 to 20, and sub--aperture size from 4.1 cm to 4.7 cm. The description of the method for estimating \ac{tau} presented here is for this updated prototype \ac{SHIMM}. 

The effective wind velocity can be estimated by acquiring the power spectrum of the Zernike defocus mode of the wavefront aberration. The defocus term has been used previously by the \ac{FADE} instrument to estimate \ac{tau} by employing a small telescope with a central obstruction to produce a ring-like defocused image \citep{Tokovinin08}. Here, however, the Zernike analysis is applied to the \ac{WFS} centroid data.


The first order Zernike modes of the atmospheric aberration, representing the angle of arrival fluctuations of the starlight  or tip/tilt of the wavefront, have the largest variance. Hence, in principle, the power spectrum of the first order modes would provide the highest SNR for estimation of the \ac{tau}. However, the tip/tilt modes include the effects of telescope shake and guiding errors, which cannot be distinguished from the atmospheric contribution. Therefore, \ac{tau} is estimated from the power spectrum of the second order defocus atmospheric term, which is unaffected by telescope shake and guiding errors. Furthermore, the defocus mode has circular symmetry so that its shape is independent of the wind direction relative to the WFS, this was verified in simulation. The \ac{SHWFS} slopes are converted into Zernike coefficients via an interaction matrix describing the local gradient of each Zernike mode across each subaperture, which can be analytically computed from gamma matrices \citep{Noll76, Townson17}. The dot product of the pseudo-inverse of this matrix with measured slopes then provides the Zernike decomposition for a particular frame. 


Utilising a linear-log scale, to display a normalised power spectrum, emphasizes the sharp localization of the energy at \ac{peakf} \citep{Hogge76}. This is defined as
 \begin{equation}
\label{eq:tau_normpower_spec}
P^{\mathrm{norm}} =  \frac{f\Phi\left(f \right)}{\int \Phi \left(f \right) \, df} \, , 
\end{equation}
where $\Phi\left(f \right)$ is the power spectrum of the defocus term.
In addition, the area under the curve is equal to the total energy of the power spectrum \citep{Roddier93}. The parameter \ac{peakf} is related to the velocity by 
\begin{equation}
\label{eq:tau_kneefreq}
f_{\mathrm{peak}} = \gamma \frac{V(h)}{d} \,, 
\end{equation}
where $\gamma$ is a constant factor related to the \ac{WFS} geometry. 

In practise, the measured power spectrum will comprise contributions from multiple turbulent layers of the atmosphere, each characterized by its wind speed and optical turbulence strength. One way of determining \ac{veff} is by fitting a linear sum of several power spectra to the measured spectrum. However, a different method was developed that proved, in simulation, to be more robust for profiles with many non-distinct turbulent layers.

\begin{figure}
\centering
\includegraphics[width=1.\columnwidth]{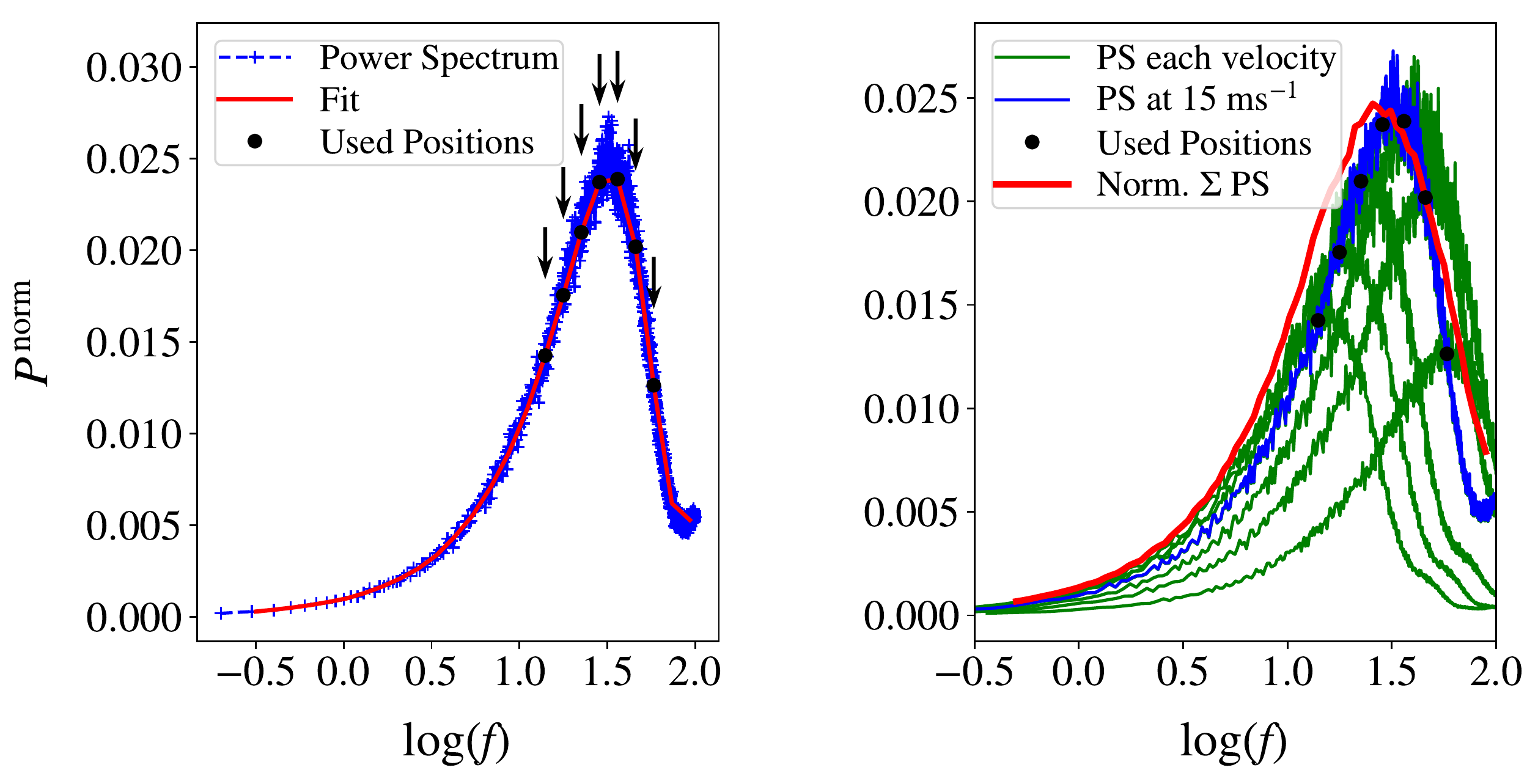}
\caption{(Left) illustrates the method for measuring the wind speed. An example spectrum of the defocus term taken from the WFS slopes (blue) from a simulated single turbulent layer travelling at 15~m~s$^{-1}$. The red line is the smoothed version of the data and the black markers indicate the sampling frequencies used for estimating the wind speed. (Right) demonstrates how this method can be used, by assuming each marker represents a turbulent layer of particular strength and speed (green spectra). The sum of these individual spectra will form a spectrum (red) similar to the original spectrum (blue). \label{fig:tau_1layer_spec}}
\end{figure}

The method adopted here is illustrated in figure \ref{fig:tau_1layer_spec}. 
First, the data is smoothed (depicted by the red line) by using a moving average of the power spectrum to reduce the scatter. Since there are fewer samples at low frequencies the moving average is completed in two parts for frequencies above and below log(0.5), where each is averaged over a different number of samples. An interpolation was used to obtain points at equal intervals (for example those denoted by the black arrow markers in the left subplot of figure \ref{fig:tau_1layer_spec}). Each marker assumes a power spectrum of particular strength and speed. Summing these individual spectra forms a power spectrum similar to that of the original power spectrum. A value for \ac{veff} is then found via the weighted sum of these contributions, using equation \ref{eq:veff}. 

\begin{figure}
\centering
\includegraphics[width=1.\columnwidth]{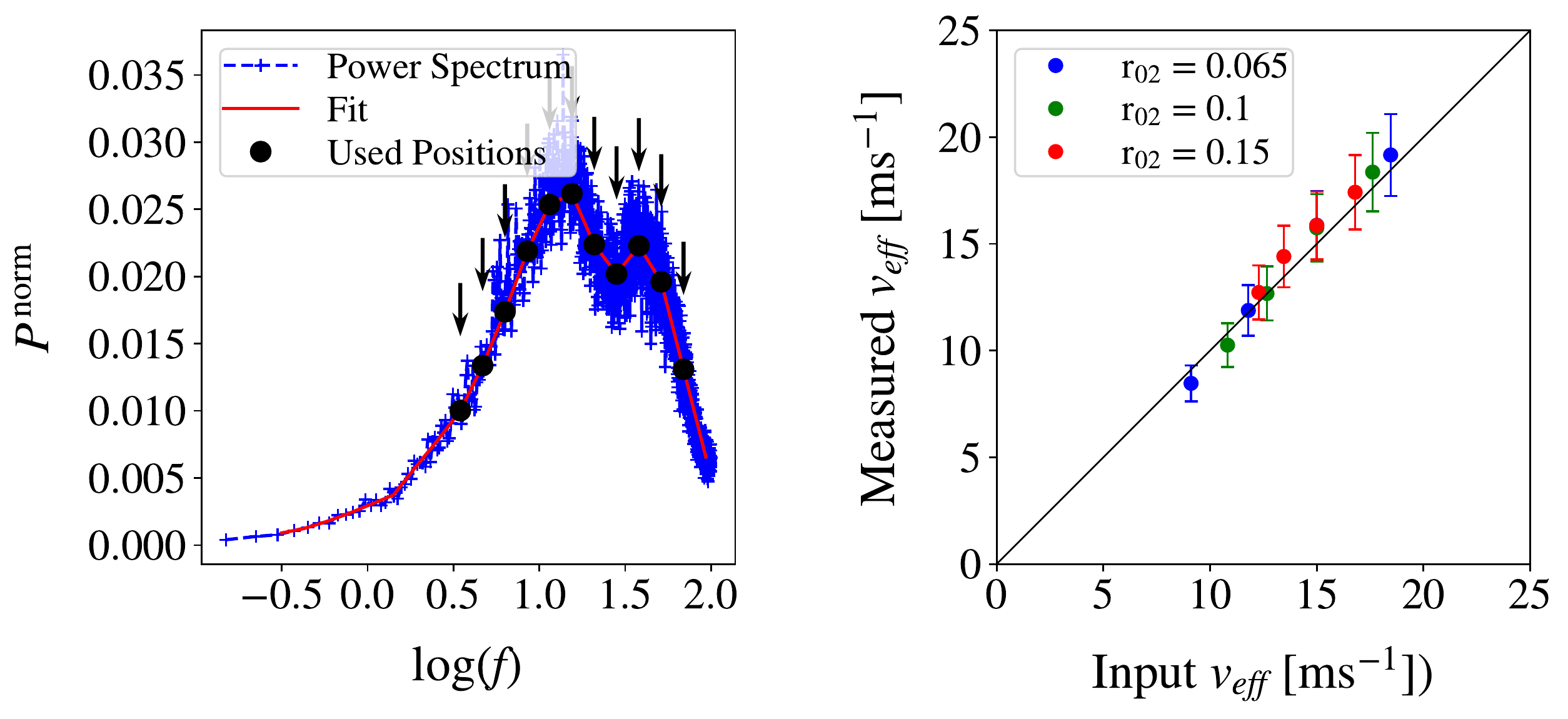}
\caption{(Left) illustrates the method for measuring the wind speed for an example spectrum of the defocus Zernike term (blue) from a simulated two turbulent layer profile travelling at 5 and 15~m~s$^{-1}$, of equal strengths. The red line is the fit and the black markers indicate the sampled frequencies used for estimating the wind speed. (Right) shows simulated results for estimating the wind speed of a two turbulent layer profile. One layer is travelling at 15~m~s$^{-1}$ with a $r_{0}$ value of 0.1 m. The second layer was travelling at a range of values between 5 to 20~m~s$^{-1}$ with \ac{r0} values of 0.065~m (blue), 0.1~m (green) and 0.15~m (red). The black line indicates x~=~y.} \label{fig:tau_2layer_simresults} 
\end{figure}

Optical turbulence profiles with more than one layer will result in the superposition of multiple peaks, each corresponding to a different layer. Figure \ref{fig:tau_2layer_simresults} (left) shows a simulated example of this, for a two layer profile with distinct turbulent layers. Figure \ref{fig:tau_2layer_simresults} (right) shows the results when applying the same method described above but with two layer turbulence profiles. The error in these values is due to the uncertainty in the interpolation given by a weighted sum of the squared residuals of the spline approximation. The simulations indicate that this method works for velocity profiles with wind speeds less than 25~m~s$^{-1}$. For the 200~Hz frame rate of the Prosillica camera and 11~inch aperture telescope, only wind speeds up to 25~m~s$^{-1}$ can be estimated. For higher wind speeds the \ac{peakf} can no longer be sampled, and only the left side (low frequency) gradient of the peak is observed. In order to measure faster layers either a detector with a faster frame rate is required or a larger telescope aperture. However, results from Stereo--SCIDAR, at La Palma in June and October 2015, show that turbulent layers with wind speeds greater than 25~m~s$^{-1}$ typically occur $\sim$17\% of the time.

\section{Field Test Results}\label{sec:field-test-results}

In this section, results from field tests are presented. It should be noted that results from both the prototype \ac{SHIMM} and the updated \ac{SHIMM} will be presented. The prototype observations took place in June and October 2015, for which there is contemporaneous \ac{Stereo--SCIDAR} data. This data is presented in section \ref{section:onsky-profiling} and \ref{section:onsky-magnitude}. However, the updated \ac{SHIMM} observations took place in July and September 2016, where there was no other profiling instrument to compare to. This data is presented in section \ref{section:onsky-tau0}. 

\subsection{Profiling Results and Corrected Fried Parameter ($r_{0}$)}
\label{section:onsky-profiling}
The results presented here were from observations that took place on the 27th - 29th June 2015 and 5th October 2015 on the roof of the \ac{INT} building, La Palma, with the prototype \ac{SHIMM}.

\begin{figure*}
\begin{subfigure}{1.\columnwidth}
  \centering
  \includegraphics[width=1.\columnwidth]{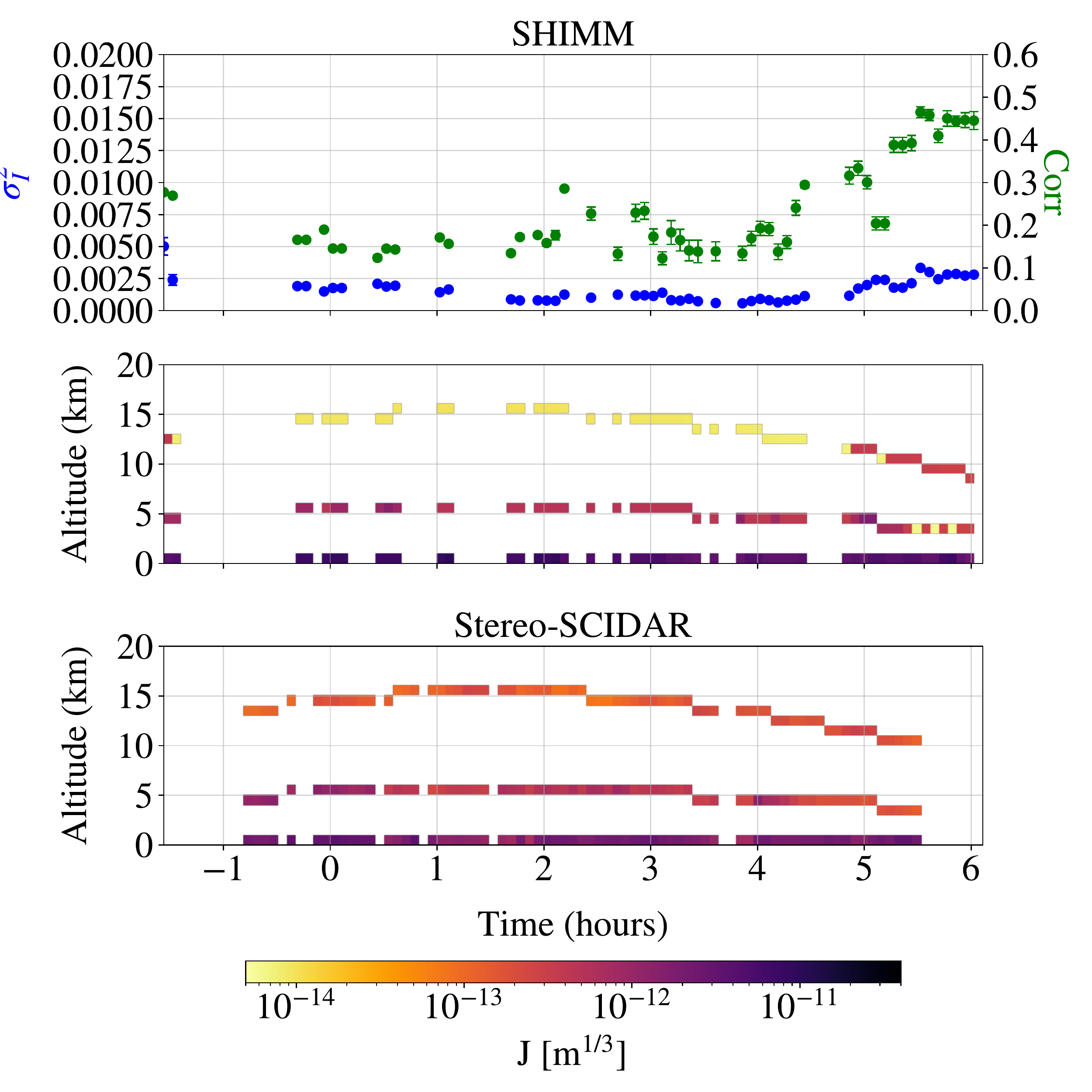}  
  \caption{27th June 2015.}
  \label{fig:sub-first}
\end{subfigure}
\begin{subfigure}{1.\columnwidth}
  \centering
  \includegraphics[width=1.\columnwidth]{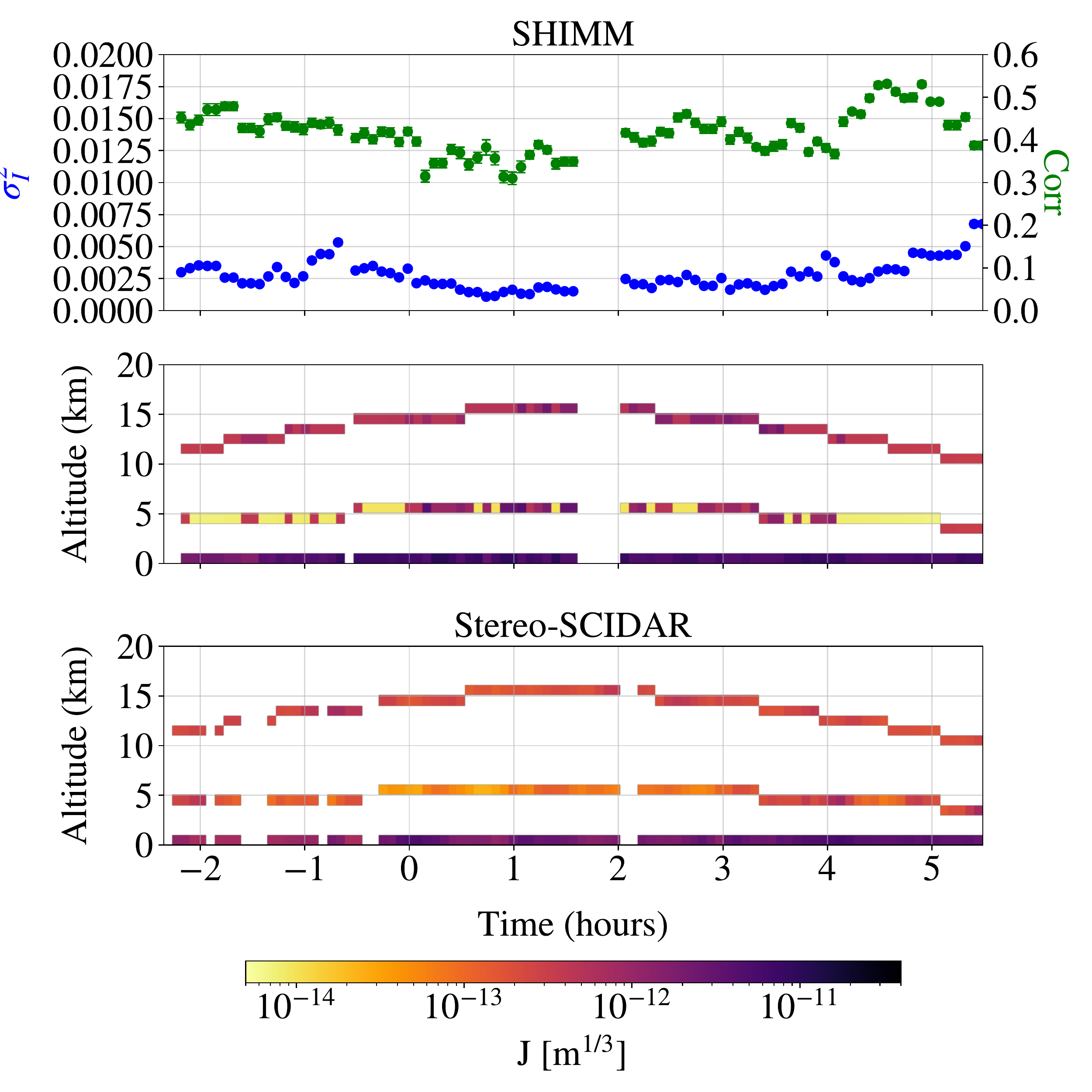}  
  \caption{28th June 2015.}
  \label{fig:sub-second}
\end{subfigure}

\vspace{1cm}
\begin{subfigure}{1.\columnwidth}
  \centering
  \includegraphics[width=1.\columnwidth]{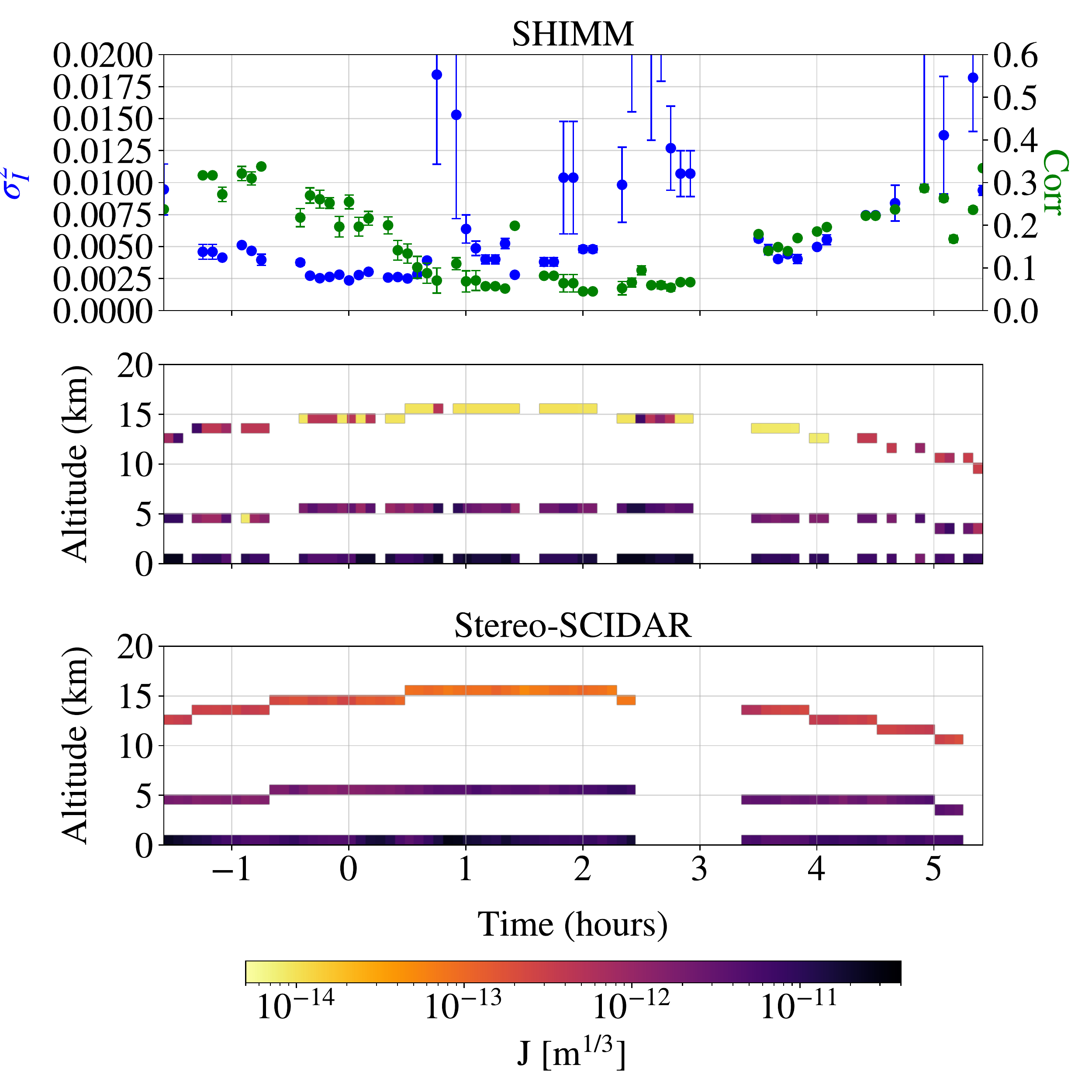}  
  \caption{29th June 2015.}
  \label{fig:sub-third}
\end{subfigure}
\begin{subfigure}{1.\columnwidth}
  \centering
  \includegraphics[width=1.\columnwidth]{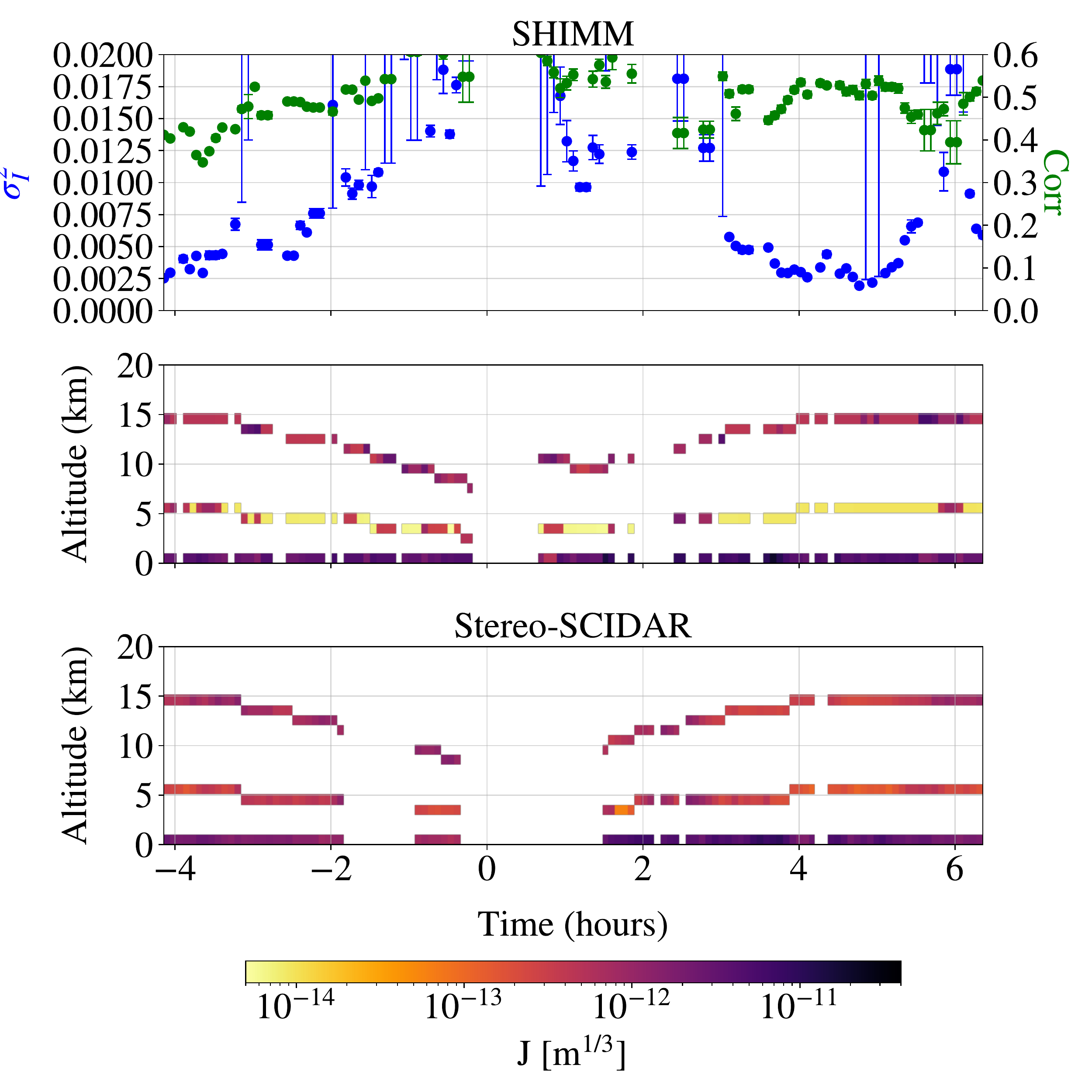}
  \caption{5th October 2015.}
  \label{fig:sub-fourth}
\end{subfigure}
\caption{Turbulence profile results for SHIMM field tests. Each panel shows (top) the correlation of the scintillation and scintillation index over the night (middle) the three--layer SHIMM profile and (bottom) the Stereo--SCIDAR profile binned to match the SHIMM binning.}
\label{fig:onsky-profiles}
\end{figure*}

Figure \ref{fig:onsky-profiles} shows the measured profile sequences for each night of observation, demonstrating how the estimated \ac{OTP} evolves with the measured scintillation parameters ($\sigma_{I}^{2}$ and Corr). The turbulence profiles have been corrected for airmass. Intervals in the profile sequences were due to target changes, telescope autoguiding corrections and occasional overcast. 

As expected, larger values of $\sigma_{I}^{2}$ are reflected by increased turbulence strength in one or both of the 5~km and 15~km layers. When the correlation of intensity fluctuations between neighbouring sub--apertures is relatively low, e.g. for most of the night of 27th June 2015 and 29th June 2015, the estimated strength of the 5~km is larger than for the 15~km layer. This situation is reversed on the nights of 28th June 2015 and 5th October 2015, when the correlation is much higher. 


Figure \ref{fig:onsky-profiles} also includes concurrent Stereo--SCIDAR profiles that have been binned in altitude to match the vertical profile resolution of the SHIMM. The Stereo--SCIDAR profiles broadly match those from the SHIMM and show similar trends. For example, figures \ref{fig:onsky-profiles} (a) and (c) show a relatively weak high altitude layer measured by both instruments, whereas figure \ref{fig:onsky-profiles} (d) shows a relatively weak intermediate altitude layer.

\begin{figure}
\centering
\includegraphics[width=1.\columnwidth]{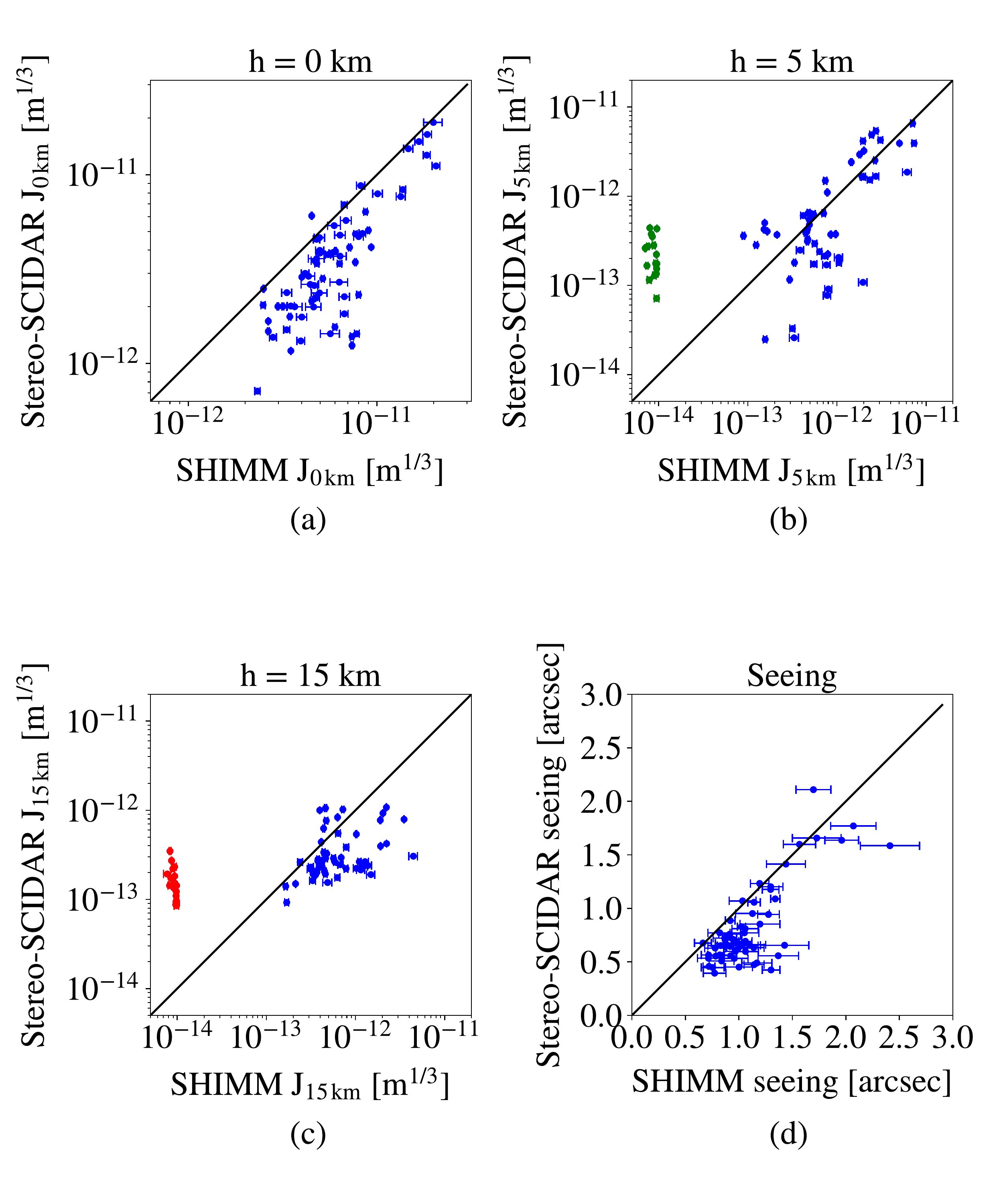}
\caption{On-sky comparison results, between the Stereo--SCIDAR mounted on the INT and the SHIMM, for the turbulence strength at 0~km (a), 5~km (b) and 15~km (c) when airmass corrected, and the total integrated seeing after scintillation correct (d). The green and red points highlight the outliers for the 5~km and 15~km respectively. \label{fig:prof_loglog} }
\end{figure}

Figure \ref{fig:prof_loglog} compares the turbulence strength at 0, 5 and 15 km (after airmass correction) between the SHIMM and Stereo--SCIDAR. The estimated strength of the ground--level turbulence is clearly correlated between the two instruments, but the SHIMM systematically measures stronger turbulence than Stereo--SCIDAR for this altitude. Figures \ref{fig:prof_loglog} (b) and (c) show closer agreement for the higher altitude (5~km and 15~km) layers, but with substantial scatter, and some bias to overestimate the turbulence at the 15 km layer. 

Figure \ref{fig:prof_loglog} (d) compares the overall seeing angle values for the SHIMM and Stereo--SCIDAR. In most cases, the SHIMM seeing value is higher than for Stereo--SCIDAR, largely as a result of the excess turbulence strength measured by the SHIMM at ground level. The instrument response function (figure \ref{fig:layer_placement}) shows that a small percentage of higher altitude turbulence may be incorrectly allocated to the ground layer, leading to an overestimation of this layer. However, since the turbulence at higher altitudes is typically weaker, this effect is likely to be small. Also if this was significant it would result in an under--correction of the seeing. It is likely that this measured excess results from local turbulence at the site of the SHIMM that did not affect the Stereo--SCIDAR measurements. The SHIMM was mounted on the roof of the INT building. Local turbulence may have been generated by heating and local wind--shear effects caused by the building itself, or turbulence within the closed telescope tube. 

There are a number of instances where the SHIMM estimates extremely weak turbulence in one of the higher altitude layers. These appear as groups of outlying points in figures \ref{fig:prof_loglog} (b) and (c) and are marked in green and red respectively for reference. We believe that these result from  two effects in the model fit to SHIMM data, as follows. 
 
When the turbulence at high altitudes is strong, both $\sigma_{I}^{2}$ and Corr are large. In this scenario, the SHIMM model correctly allocates strong turbulence to the 15~km layer. However, the value of the turbulence strength in the intermediate 5~km layer can become poorly constrained in these circumstances, i.e. its value no longer has a large effect on the scintillation parameters. The distribution of turbulence strength between the 0~km and 5~km layers is then more poorly constrained and noisy. In many cases, the SHIMM model fit chooses the minimum allowed value for the 5~km turbulence strength, producing the cluster of points shown in green in figure \ref{fig:prof_loglog} (b). This effect can also be seen in figures  \ref{fig:onsky-profiles} (b) and (d) for the nights of 28th June 2015 and 5th October 2015, when the 15~km turbulence is consistently strong. The SHIMM turbulence strength for the 5~km layer is very variable and is not matched by similar variability of the Stereo--SCIDAR measurements. Although this is a limitation of the SHIMM method for profile estimation, we note that in these circumstances and according to equation \ref{eq:heff}, the high altitude (15~km layer) will dominate the estimate of \ac{theta} from SHIMM as well as the correction of r$_{0}$ for scintillation effects. The increased uncertainty of the 5~km layer strength is then relatively unimportant. 

\begin{figure}
\centering
\includegraphics[width=.7\columnwidth]{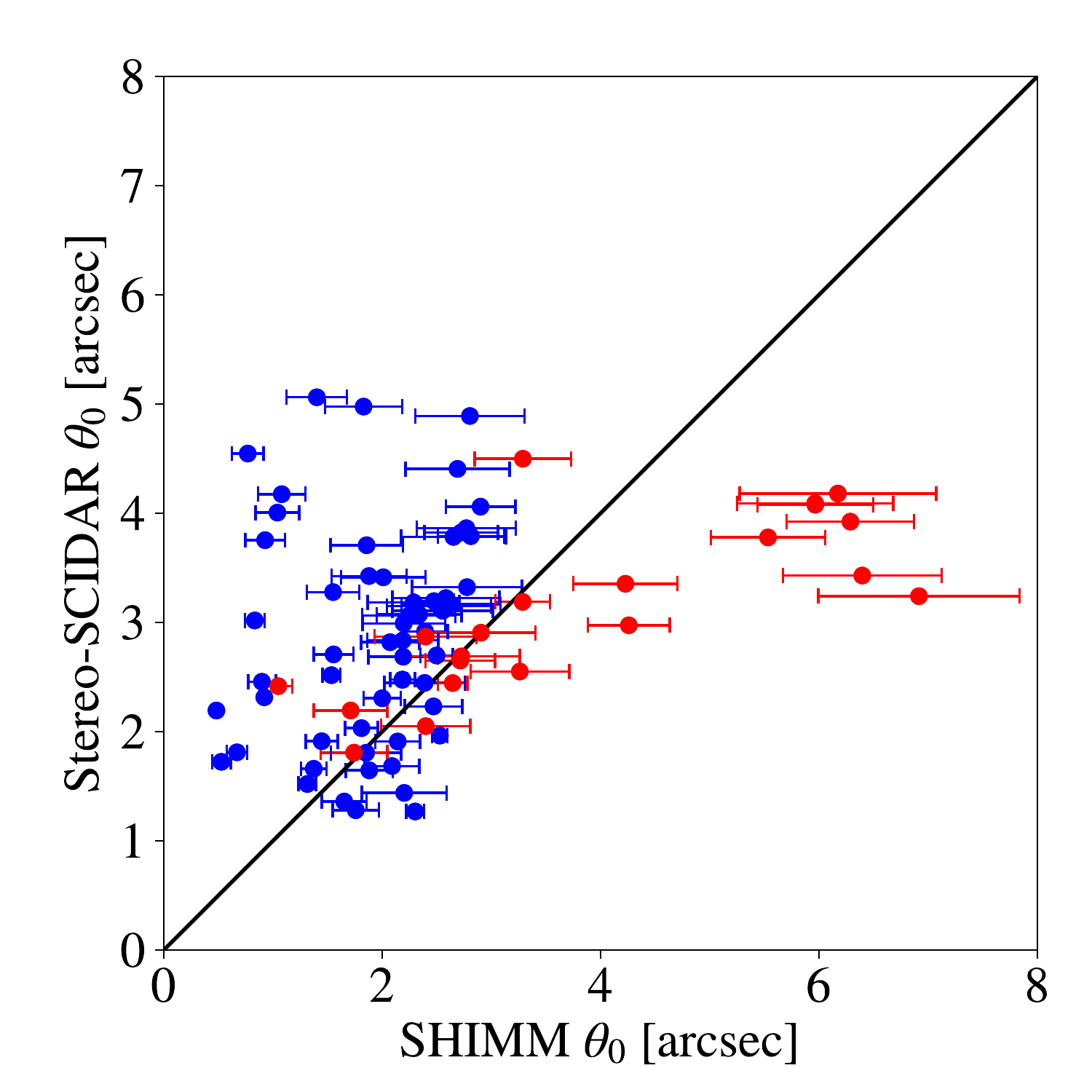}
\caption{Comparison of the isoplanatic angle measured by Stereo--SCIDAR mounted on the INT and the SHIMM located on the roof of the INT. The SHIMM estimate is derived from the three--layer profiles (shown in figure \ref{fig:prof_loglog}), whereas the Stereo--SCIDAR estimate is derived from its full high-resolution profile. The red points refer to periods when the highest altitude turbulence was very weak. \label{fig:prof_theta0} }
\end{figure}

Conversely, when the turbulence at high altitudes is very weak, both the scintillation index and correlation values become small. In these circumstances, the turbulence strength in the 15~km layer was often underestimated by the prototype SHIMM, in particular on the night of 27th June 2015. In a number of cases, the SHIMM analysis chooses the minimum allowed value for the 15~km turbulence strength, producing the cluster of outlying points shown in red in figure \ref{fig:prof_loglog} (c). We believe that this most likely results from the increased sensitivity of the technique to the effects of shot noise and, in particular, the readout noise of the detector when the scintillation is very weak. For example, higher readout noise results in some underestimation of Corr, and hence the strength of the 15~km layer in the model fit. We expect that this issue will be greatly reduced for more recent camera models, in particular new CMOS cameras capable of delivering high frame rates with very low readout noise. This error in the estimation of the high layer in weak scintillation has minimal impact on the correction of r$_{0}$ for scintillation effects. It is more significant for the estimation of \ac{theta}, resulting in an overestimation of its value, as indicated in figure in \ref{fig:prof_theta0}. 

Figure \ref{fig:prof_theta0} compares the estimated \ac{theta} from the low-resolution profile of the \ac{SHIMM} to that estimated from the high-resolution profile of \ac{Stereo-SCIDAR}.  Figure \ref{fig:prof_theta0} additionally shows instances where the \ac{SHIMM} underestimates \ac{theta}, due to the overestimation of the turbulence strength at the high-altitude layer. This slight bias may be due to the response of the model used, as shown in figure \ref{fig:layer_placement}. \ac{Stereo-SCIDAR} is a much higher resolution instrument and therefore will determine the strength of individual turbulent layers. The \ac{SHIMM} however, will split the placement of the turbulence of the layers between the mid and high-altitude. 
 
\subsection{Instrument repeatability and limiting magnitude}
\label{section:onsky-magnitude}
\begin{figure}
	\includegraphics[width=1.\columnwidth]{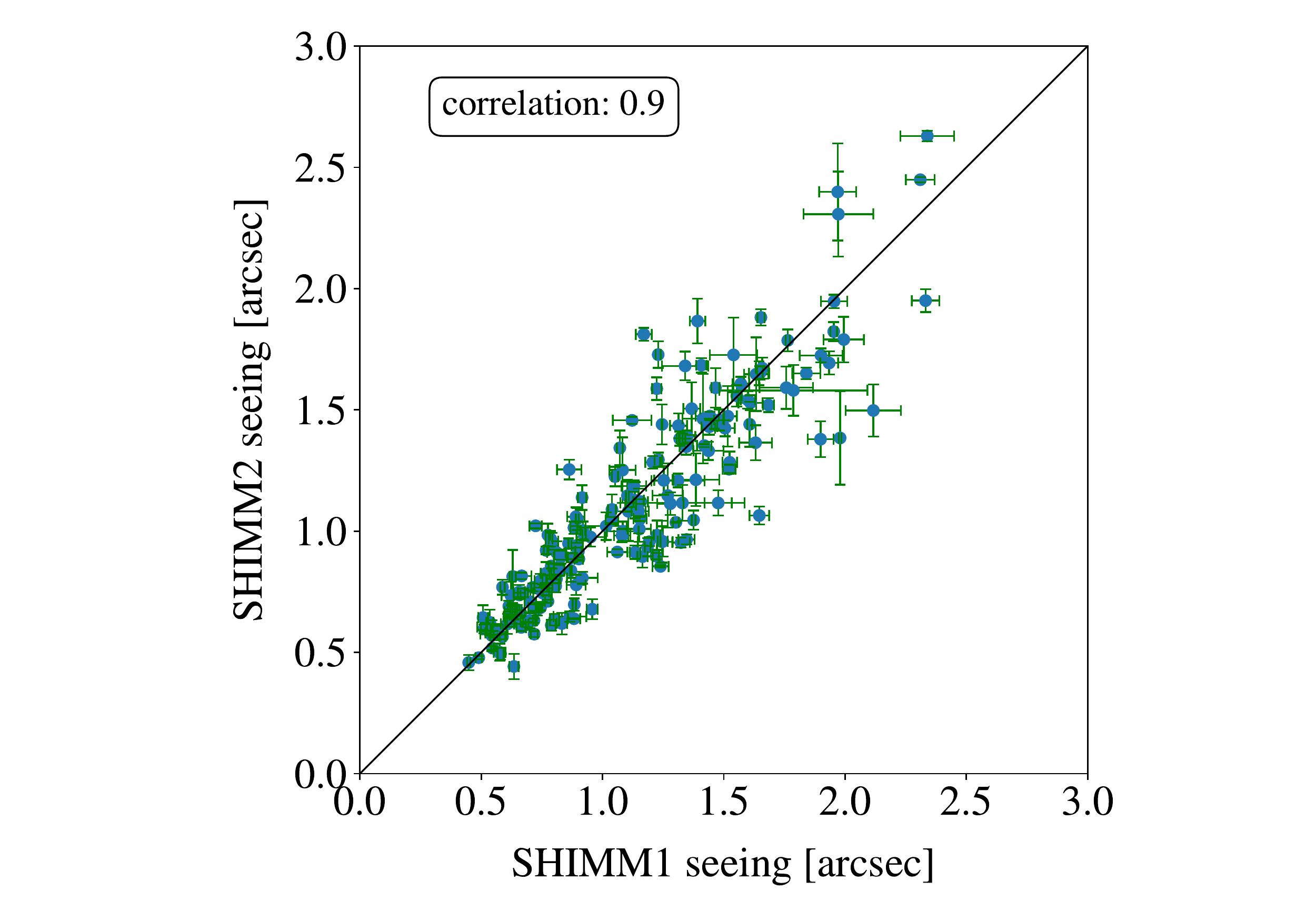}
    \caption{Comparison of contemporaneous measurements of the seeing angle from two identical SHIMM instruments operating side by side and observing the same target star, over several nights in June and October 2015. Taken from \citet{Perera16}.}\label{fig:both}
\end{figure}

Observations were made simultaneously with two identical prototype \ac{SHIMM} instruments located side-by-side on the roof of the \ac{INT} building, between 25th - 30th June 2015 and 30th September - 5th October 2015. The results are shown in figure \ref{fig:both}, indicating very good agreement between the two \ac{SHIMM}s when they observe the same target. Some scatter in the values is expected due to statistical noise, since they observe along slightly different lines of sight to the target star. In particular, the effects of any local turbulence would be slightly different for the two instruments and would be slow to converge in the measurements. However, there is no significant bias in the measurements over a large range of seeing conditions, indicating very good repeatability in the construction and calibrations of the instrument.  

\begin{figure}
\includegraphics[width=1.\columnwidth]{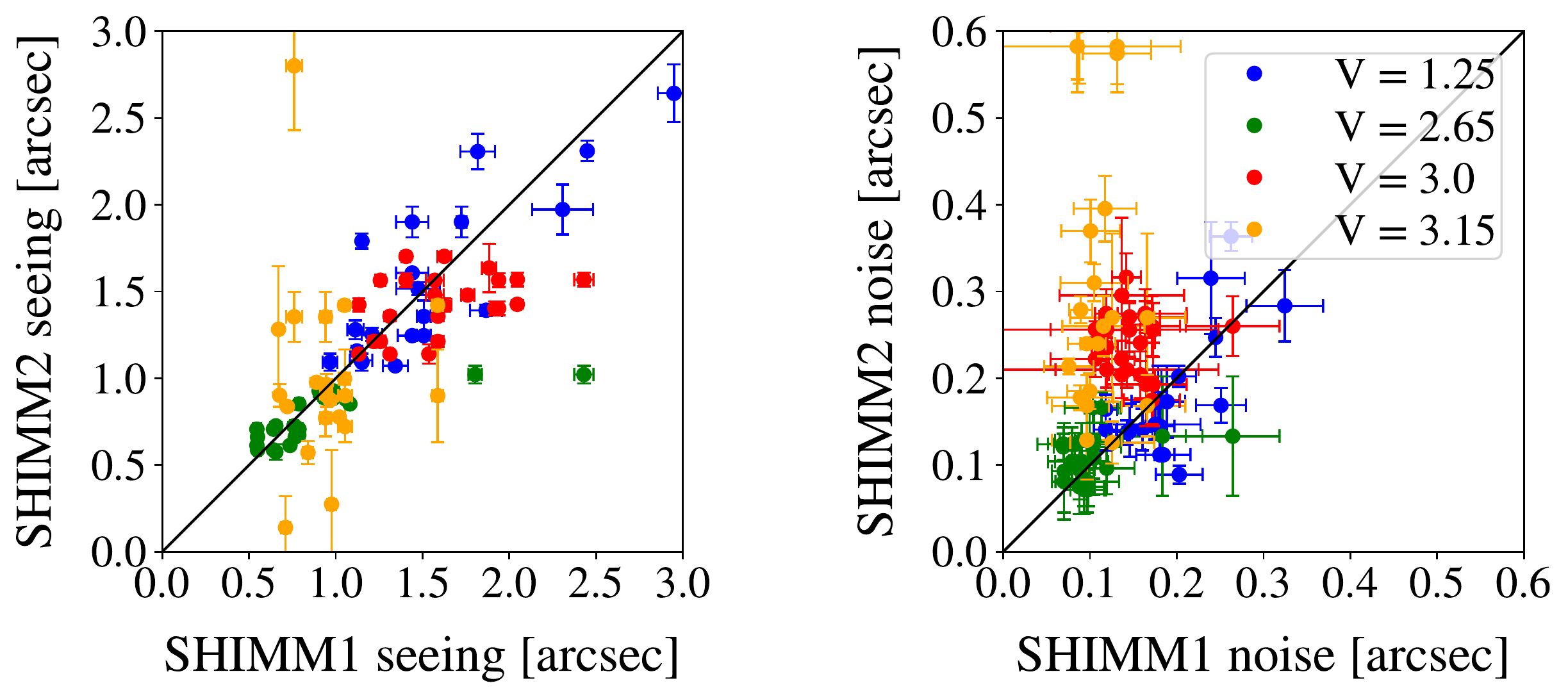}
\caption{Measured seeing (left) and noise (right) comparisons between the two SHIMM instruments, where SHIMM1 observed targets of V = 0.03 or brighter and SHIMM2 observed a range of target magnitudes.}\label{fig:shimm_maglim}
\end{figure}

The arrangement of two identical SHIMMs (SHIMM1 and SHIMM2) operating simultaneously also permitted direct testing of the limiting target magnitude of the instrument. SHIMM1 observed a bright reference star Vega (V~=~0.03), whereas SHIMM2 observed fainter stars at similar zenith angles and positions as the bright reference star. Figure \ref{fig:shimm_maglim} shows the results of the seeing and the noise estimated by the two SHIMM instruments when observing targets of different brightness. It can be seen that there is a good correlation between the measurements for target magnitudes V $<$ 3.  The overall noise for SHIMM2 diverges significantly from that of SHIMM1 for targets fainter than V = 2.65. This indicates that the noise for fainter targets is dominated by shot noise, rather than statistical noise. Hence we conclude that for this SHIMM configuration, although third magnitude target stars can be employed, the preference is to use second magnitude stars or brighter whenever possible.

\subsection{Coherence Time ($\tau_{0}$) Measurements}
\label{section:onsky-tau0}
Observations were made in July and September 2016 on the roof of the \ac{WHT} building in La Palma, with the upgraded \ac{SHIMM}. For these dates there were no concurrent Stereo--SCIDAR observations, therefore the statistics between the two instruments are compared instead.


\begin{figure}
	\includegraphics[width=1.\columnwidth]{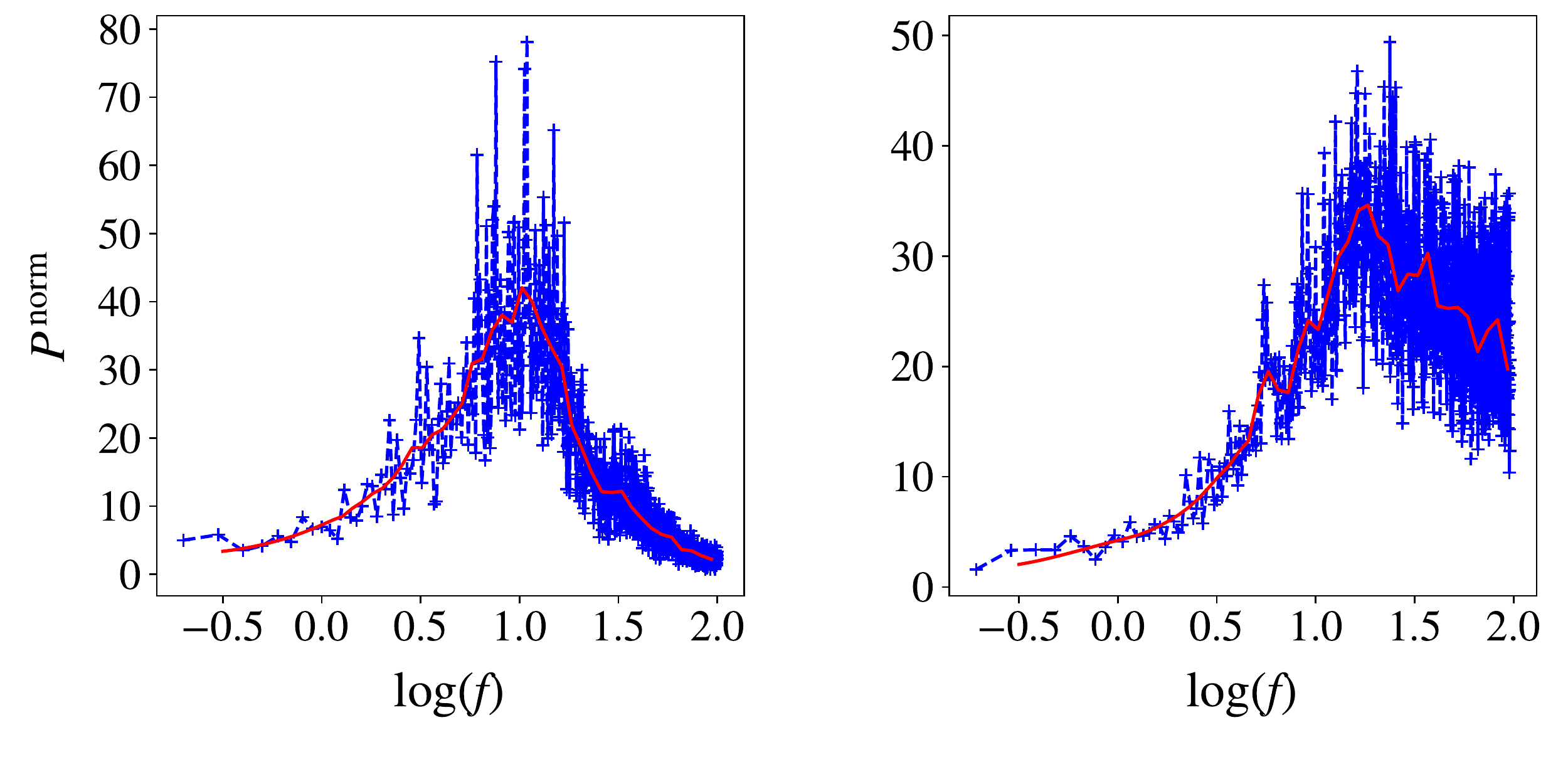}
    \caption{Example power spectra of the defocus term of on-sky data taken with the C11-SHIMM at the INT site, La Palma. The blue line depicts the raw power spectra data, and the red line displays the interpolated fit used to calculate the \ac{veff}. The estimated \ac{veff} is $4.7 \pm 0.2$ and $17.0 \pm 0.5 $~m~s$^{-1}$.}\label{fig:shimm_tau0_realexamples}
\end{figure}

\begin{figure}
	\includegraphics[width=1\columnwidth]{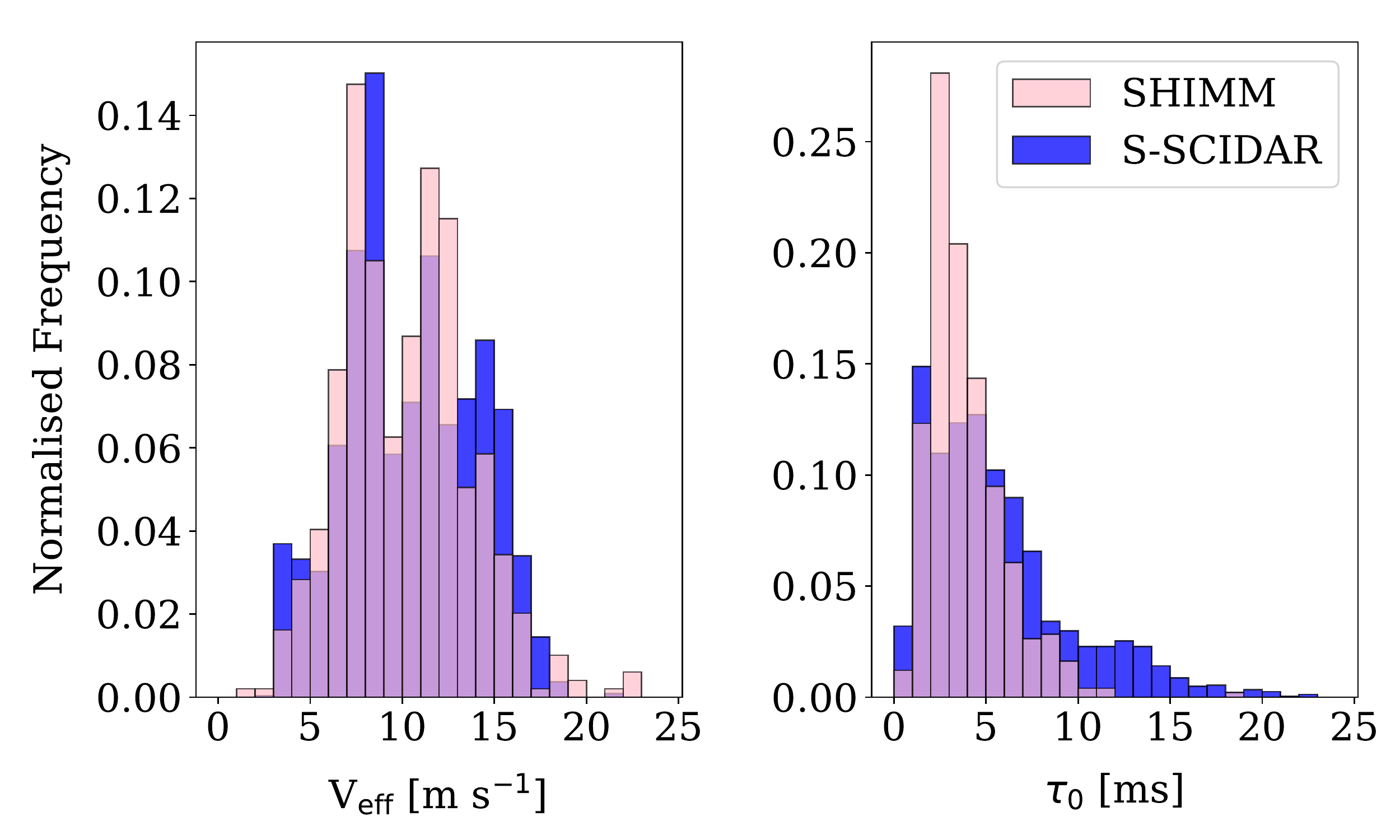} 
    \caption{Histogram of veff (left) and \ac{tau} (right) measured by the Stereo--SCIDAR  in June and October 2015 over $\sim$2500 data points (blue) and SHIMM in July and September 2016 over $\sim$500 data points (pink) taken from two different observing sites mounted on the INT and roof of the WHT at La Palma respectively.}\label{fig:hist_tau0}
\end{figure}
Figure \ref{fig:shimm_tau0_realexamples} shows example power spectra of the defocus Zernike term measured with the modified SHIMM. 
Figure \ref{fig:hist_tau0} illustrates the normalised frequencies of \ac{veff} and \ac{tau} estimated by the SHIMM on La Palma in 2016, as well as those for the Stereo--SCIDAR over all its observations during July and October 2015.  The median values of \ac{veff} for the SHIMM and Stereo--SCIDAR are 10.21 $\pm$ 0.16~m~s$^{-1}$ and 10.44 $\pm$ 0.07~m~s$^{-1}$ respectively, with standard error. This shows good agreement in the statistical distribution of \ac{veff}, indicating that the method described in section \ref{section:tau0} can be used to estimate this parameter accurately. The median value of \ac{tau} for the SHIMM and Stereo--SCIDAR is 3.88 $\pm$ 0.09~ms and 5.62 $\pm$ 0.08~ms respectively, indicating that SHIMM underestimates the coherence time. 
Since \ac{veff} is in good agreement, according to equation \ref{eq:tau0}, this arises because the SHIMM underestimates the value of \ac{r0} relative to Stereo--SCIDAR. This is likely a result of local ground--level seeing effects at the site of the SHIMM. 


\section{Discussion \& Conclusions}

We have presented the concept, design and preliminary field test results for the SHIMM -- a versatile, low-cost, portable seeing monitor. The instrument was built from off-the-shelf components making it easy to duplicate, and therefore ideal for investigating differences in the seeing around large observing sites. The improvements provided by the SHIMM over the traditional DIMM seeing monitor were presented, including (i) estimation of the seeing angle independent of noise bias; (ii) seeing estimates corrected for bias by scintillation effects; (iii) a three--layer optical turbulence profile and (iv) estimation of atmospheric coherence time. The methodology and results from numerical simulation were presented for each of these features. 

Results from field tests, obtained at Roque de los Muchachos Observatory, La Palma, showed a generally good correlation with contemporaneous Stereo--SCIDAR measurements of the optical turbulence strength at low, intermediate and high altitudes, showing similar trends throughout the observed nights. Although the SHIMM generally overestimated the overall value of the seeing angle relative to Stereo--SCIDAR, this 
resulted largely from excess ground--level turbulence local to the \ac{SHIMM}.

Some limitations of the SHIMM method for the estimation of the turbulence profile were evidenced. In conditions of strong scintillation, the distinction of turbulence strength between low and intermediate altitudes is poorly constrained, however, this has minimal impact on the estimation of \ac{r0} and \ac{theta}. High altitude turbulence strength was underestimated during conditions of very weak scintillation, leading to overestimation of \ac{theta}. This likely resulted from increased sensitivity to the effects of detector noise in these conditions, therefore the application of new low--noise detectors will be desirable for future implementations of the \ac{SHIMM}. At times the high altitude layer overestimates the turbulence compared to Stereo--SCIDAR resulting in underestimating \ac{theta}. This is likely due to the differences in the responses of the instruments, and the fact that Stereo--SCIDAR produces a high--resolution profile.


The robustness and magnitude limitation were tested by using two identical \ac{SHIMM}s. The results showed a good correlation between the two instruments, indicating that the \ac{SHIMM} is a reliable and easy-to-duplicate instrument. It was determined that, although targets of magnitude V~$<$~2 are preferable due to the reduced noise, targets as faint as V~$=$~3 could still recover the same estimate for the seeing angle, resulting in full sky coverage. To reduce statistical noise, data sets can be averaged over 1 - 5 minutes. Averaging over longer periods will not produce accurate results since the atmospheric turbulence profile is likely to evolve significantly over this period.


An upgraded version of the SHIMM was later implemented to permit estimation of \ac{tau}. Concurrent measurements of \ac{tau} from the SHIMM and another profiling instrument was not possible. However, the statistical distribution of \ac{veff} measured by the SHIMM was consistent with that for the Stereo--SCIDAR measured over different nights. Differences in the distributions of \ac{tau} values for the SHIMM and Stereo--SCIDAR is suspected to have resulted largely from the excess ground level turbulence strength at the SHIMM site.  This indicates that the SHIMM can  estimate \ac{veff} and \ac{tau} accurately from the power spectrum of the defocus term measured by the \ac{SHWFS}. 

More concurrent data from the SHIMM and Stereo--SCIDAR is required to determine if there is a limitation in measuring turbulence profiles accurately under a range of atmospheric conditions.



\section*{Acknowledgements}
This work was supported by the Science and Technology Facilities Council [ST/K501979/1]. We would like to thank the Isaac Newton Group of Telescopes for access to the INT. The INT is operated on the island of La Palma by the Isaac Newton Group in the Spanish Observatorio del Roque de los Muchachos of the Instituto de Astrofsica de Canarias. James Osborn acknowledges the UK Research and Innovation Future Leaders Fellowship (MR/S035338/1). The authors would also like to thank the reviewer for their comments and efforts towards improving this manuscript.

\section*{Data Availability}
The data in this underlying article was provided by the SHIMM and Stereo--SCIDAR consortium by permission. Data will be shared on request to the corresponding author with the permission of the SHIMM and Stereo--SCIDAR consortium.



\bibliographystyle{mnras}
\bibliography{example} 

\begin{thebibliography}{}
\makeatletter
\relax
\def\mn@urlcharsother{\let\do\@makeother \do\$\do\&\do\#\do\^\do\_\do\%\do\~}
\def\mn@doi{\begingroup\mn@urlcharsother \@ifnextchar [ {\mn@doi@}
  {\mn@doi@[]}}
\def\mn@doi@[#1]#2{\def\@tempa{#1}\ifx\@tempa\@empty \href
  {http://dx.doi.org/#2} {doi:#2}\else \href {http://dx.doi.org/#2} {#1}\fi
  \endgroup}
\def\mn@eprint#1#2{\mn@eprint@#1:#2::\@nil}
\def\mn@eprint@arXiv#1{\href {http://arxiv.org/abs/#1} {{\tt arXiv:#1}}}
\def\mn@eprint@dblp#1{\href {http://dblp.uni-trier.de/rec/bibtex/#1.xml}
  {dblp:#1}}
\def\mn@eprint@#1:#2:#3:#4\@nil{\def\@tempa {#1}\def\@tempb {#2}\def\@tempc
  {#3}\ifx \@tempc \@empty \let \@tempc \@tempb \let \@tempb \@tempa \fi \ifx
  \@tempb \@empty \def\@tempb {arXiv}\fi \@ifundefined
  {mn@eprint@\@tempb}{\@tempb:\@tempc}{\expandafter \expandafter \csname
  mn@eprint@\@tempb\endcsname \expandafter{\@tempc}}}

\bibitem[\protect\citeauthoryear{Aristidi, Fanteï-Caujolle, Ziad, Dimur,
  Chabé  \& Roland}{Aristidi et~al.}{2014}]{Aristidi14}
Aristidi E.,  Fanteï-Caujolle Y.,  Ziad A.,  Dimur C.,  Chabé J.,   Roland
  B.,  2014, {A new generalized differential image motion monitor},
  \mn@doi{10.1117/12.2056201}, \url {http://dx.doi.org/10.1117/12.2056201}

\bibitem[\protect\citeauthoryear{Bally, Theil, Billawalla, Potter, Loewenstein,
  Mrozek  \& Lloyd}{Bally et~al.}{1996}]{Bally96}
Bally J.,  Theil D.,  Billawalla Y.,  Potter D.,  Loewenstein R.,  Mrozek F.,
  Lloyd J.~P.,  1996, PASA, 13, 22

\bibitem[\protect\citeauthoryear{Butterley, Wilson  \& Sarazin}{Butterley
  et~al.}{2006}]{Butterley06}
Butterley T.,  Wilson R.~W.,   Sarazin M.,  2006, MNRAS, 369, 835

\bibitem[\protect\citeauthoryear{Fohring}{Fohring}{2014}]{FohringThesis}
Fohring D.,  2014, Phd thesis, University of Durham

\bibitem[\protect\citeauthoryear{Goodwin, Jenkins  \& Lambert}{Goodwin
  et~al.}{2007}]{Goodwin07}
Goodwin M.,  Jenkins C.,   Lambert A.,  2007, \mn@doi [Opt. Express]
  {10.1364/OE.15.014844}, 15, 14844

\bibitem[\protect\citeauthoryear{Guesalaga, Perera, Osborn, Sarazin, Neichel
  \& Wilson}{Guesalaga et~al.}{2016}]{Guesalaga16}
Guesalaga A.,  Perera S.,  Osborn J.,  Sarazin M.,  Neichel B.,   Wilson R.~W.,
   2016, in TBC ed.,  Proc. SPIE Vol. 9909, Astronomical Telescopes and
  Instrumentation. p. In these proceedings

\bibitem[\protect\citeauthoryear{Hardy}{Hardy}{1998}]{Hardy98}
Hardy J.~W.,  1998, Adaptive Optics for Astronomical Telescopes.
Oxford University Press, New York

\bibitem[\protect\citeauthoryear{Hogge \& Butts}{Hogge \&
  Butts}{1976}]{Hogge76}
Hogge C.~B.,  Butts R.~R.,  1976, IEEE Trans. Antennas Propag., 24, 144

\bibitem[\protect\citeauthoryear{Noll}{Noll}{1976}]{Noll76}
Noll R.~J.,  1976, \mn@doi [J. Opt. Soc. Am.] {10.1364/JOSA.66.000207}, 66, 207

\bibitem[\protect\citeauthoryear{O'Donovan, Young, Warner, Buscher, Wilson,
  Boysen, Seneta  \& Keen}{O'Donovan et~al.}{2003}]{Donovan03}
O'Donovan B.,  Young J.~S.,  Warner P.~J.,  Buscher D.~F.,  Wilson D. M.~A.,
  Boysen R.~C.,  Seneta E.~B.,   Keen J.,  2003, {DIMMWIT: comparing
  atmospheric seeing values measured by a differential image motion monitor,
  which is transportable, and COAST}, \mn@doi{10.1117/12.459761}, \url
  {http://dx.doi.org/10.1117/12.459761}

\bibitem[\protect\citeauthoryear{Osborn, Wilson, Dhillon, Avila  \&
  Love}{Osborn et~al.}{2011}]{Osborn11}
Osborn J.,  Wilson R.~W.,  Dhillon V.~S.,  Avila R.,   Love G.~D.,  2011,
  MNRAS, 411, 1223

\bibitem[\protect\citeauthoryear{Osborn, Föhring, Dhillon  \& Wilson}{Osborn
  et~al.}{2015}]{Osborn15}
Osborn J.,  Föhring D.,  Dhillon V.~S.,   Wilson R.~W.,  2015, MNRAS, 452,
  1707

\bibitem[\protect\citeauthoryear{Perera, Wilson, Osborn  \& Butterley}{Perera
  et~al.}{2016}]{Perera16}
Perera S.,  Wilson R.~W.,  Osborn J.,   Butterley T.,  2016, in Marchetti E.,
  Close L.~M.,   V{\'e}ran J.-P.,  eds, ~ Vol. 9909, Adaptive Optics Systems V.
  SPIE, p. 99093J, \mn@doi{10.1117/12.2231680}, \url
  {https://doi.org/10.1117/12.2231680}

\bibitem[\protect\citeauthoryear{Robert, Voyez, V{\'{e}}drenne  \&
  Mugnier}{Robert et~al.}{2011}]{Robert11}
Robert C.,  Voyez J.,  V{\'{e}}drenne N.,   Mugnier L.,  2011, Observatory

\bibitem[\protect\citeauthoryear{Roddier, Northcott, Graves,   \&
  McKenna}{Roddier et~al.}{1993}]{Roddier93}
Roddier F.,  Northcott M.~J.,  Graves J.~E.,    McKenna D.~L.,  1993, J. Opt.
  Soc. Am. A, 10, 957

\bibitem[\protect\citeauthoryear{Sarazin \& Roddier}{Sarazin \&
  Roddier}{1989}]{Sarazin89}
Sarazin M.,  Roddier F.,  1989, Astron. \& Astrop., 227, 294

\bibitem[\protect\citeauthoryear{Tokovinin \& Kornilov}{Tokovinin \&
  Kornilov}{2007}]{Tokovinin07A}
Tokovinin A.,  Kornilov V.,  2007, MNRAS, 381, 1179

\bibitem[\protect\citeauthoryear{Tokovinin, Kellerer  \& Foresto}{Tokovinin
  et~al.}{2008}]{Tokovinin08}
Tokovinin A.,  Kellerer A.,   Foresto V. C.~D.,  2008, A\&A, 477, 671

\bibitem[\protect\citeauthoryear{Townson, Reeves, Osborn, Bitenc, Laidlaw  \&
  Farley}{Townson et~al.}{2017}]{Townson17}
Townson M.~J.,  Reeves A.~P.,  Osborn J.,  Bitenc U.,  Laidlaw D.~J.,   Farley
  O.,  2017, In Prep

\bibitem[\protect\citeauthoryear{Vedrenne, Michau, Robert  \& Conan}{Vedrenne
  et~al.}{2007}]{Vedrenne07}
Vedrenne N.,  Michau V.,  Robert C.,   Conan J.,  2007, Opt. Lett, 32, 2659

\bibitem[\protect\citeauthoryear{Wilson, O'Mahony, Packham  \& Azzaro}{Wilson
  et~al.}{1999}]{Wilson99}
Wilson R.~W.,  O'Mahony N.,  Packham C.,   Azzaro M.,  1999, MNRAS, 309, 379

\makeatother
\end{thebibliography}








\bsp	
\label{lastpage}
\end{document}